\documentclass[lineno]{jfm}
\usepackage{graphicx}
\usepackage{lineno}
\usepackage{epstopdf, epsfig}
\usepackage{amsmath}
\usepackage{cases} 
\usepackage{IEEEtrantools}
\usepackage[section]{placeins}
\usepackage{xcolor}

\newcommand{\e}{\mathrm{e}}
\newcommand{\ui}{\mathrm{i}}

\newcommand{\acc}[1]{\textcolor{black}{#1}}
\newcommand{\add}[1]{\textcolor{black}{#1}}
\newcommand{\addd}[1]{\textcolor{black}{#1}}
\newcommand{\adddd}[1]{\textcolor{black}{#1}}

\shorttitle{Sound due to shock and instability waves interaction}
\shortauthor{Binhong Li and Benshuai Lyu}

\title{Acoustic emission due to the interaction between shock and instability waves in 2D supersonic jet flows}

\author{B. Li\aff{1}
   and B. Lyu\aff{1}
\corresp{\email{b.lyu@pku.edu.cn}}
}

\affiliation{\aff{1}State Key Laboratory of Turbulence and Complex
    Systems, College of Engineering, Peking University,  5 Yiheyuan Road,
    Haidian District, Beijing 100871, China
}
\begin{document}

\maketitle
\begin{abstract}
An analytical model is developed to study the sound produced by the interaction between shock and instability waves in two-dimensional supersonic jet flows. The jet is considered to be of vortex-sheet type and 2D Euler equations are linearised to determine the governing equations for shock, instability waves, and their interaction. Pack's model is used to describe shock waves, while instability waves are calculated using spatial stability analysis. The interaction between shock and instability waves can be solved analytically by performing Fourier \add{transform} and subsequently using the method of steepest descent. Sound produced by the interaction between the instability wave and a single shock cell is studied first, after which that due to a number of cells follows. We find that the model developed in this study can correctly predict the frequencies of the fundamental screech tone and its first and second harmonics. We show that the predicted sound directivity, even from a single shock cell, is in good agreement with experimental data. In particular, this model shows the strongest noise emission close to the upstream direction but the emitted noise starts to rapidly decay as the observer angle approaches 180 degrees, which is in accordance with experimental results; this suggests that the \add{effective} noise from a single shock cell is far from of the monopole type as assumed in the classical Powell's model. We find that the noise directivity is very sensitive to the local growth rate of the instability waves and the noise is generated primarily through the Mach wave mechanism.
\end{abstract}
\section{Introduction}
\label{sec:intro}
In many applications such as rocket engines, imperfectly-expanded supersonic
jets are often accompanied by a powerful emission of tonal sounds. Such tones
are commonly referred to as jet screech. In addition to the intensified noise
emission, it may also lead to disastrous structure damage because of sonic
fatigue. It is, therefore, of practical importance to understand the mechanism of
supersonic jet screech and to devise effective ways to suppress these screech
tones. 

Screech occurs in imperfectly-expanded supersonic jets, and such jets are often
characterized by quasi-periodic shock cells and are complex in nature. Jet screech was first
discovered in the experiment conducted by Powell in the 1950s. In his pioneering
work,~\citet{19533Powell} proposed the well-established feedback loop
which consists of four stages, i.e. the instability growth in jets, the
interactions between shock and instability waves, the acoustic waves
propagating upstream, and the receptivity of the shear layer at the nozzle lip.
Powell proposed the phase and gain conditions which must be satisfied to sustain the feedback loop. For the phase condition, the frequency of the fluctuation $f$ is supposed to close the feedback loop as
\begin{equation}
    \frac{N}{f}=\frac{d}{U_{c}}+\frac{d}{c^{*}}+\Psi,
    \label{phase_condition}
\end{equation}
where $d$ denotes the distance between the nozzle lip and the sound source, 
$U_{c}$ is the average velocity of the instability waves travelling downstream, $c^{*}$ is the speed of sound propagating upstream,
$\Psi$ represents an additional phase delay, and $N$ is an integer. For the gain condition, the gain from each of the four stages must satisfy
\begin{equation}
    Q\eta_{s}\eta_{u}\eta_{r}\geq1,
\end{equation}
where $Q$ denotes the gain associated with the growth of the instability waves, and $\eta_{s}$, $\eta_{u}$ and $\eta_{r}$ represent the efficiencies of energy transmission in the last three stages, respectively. The resonance conditions were then reconsidered by examining the energy exchange between the instability and acoustic waves at the sound source location and the nozzle exit~\citep{NewC_3}. In the recent work, these conditions~\citep{NewC_3} have been rewritten in terms of magnitude and phase conditions, details of which can be found in~\citet{NewC_2} and~\citet{NewC_1}.

Two important characteristics of jet screech are widely studied over the past few decades. The first is the screech frequency. Considering the feedback enhancement during the
loop, Powell proposed that the screech frequency $f$ can be calculated via 
\begin{equation}
f=\frac{U_{c}}{s(1+M_{c})},
\label{equ_screech f}
\end{equation}
where $s$, $U_{c}$, and $M_{c}$ are the shock cell spacing, the convection velocity and the \adddd{convective} Mach number of the instability waves, respectively. Subsequently,~\citet{1986Tam_Proposed} proposed the weakest link theory suggesting the screech as the limit of the broadband shock-associated noise~(BBSAN) when the observer angle approaches $180^\circ$. In 1999, \citet{Panda_standingwave} discussed the link between screech and hydrodynamic-acoustic~(HA) standing waves, and a new formula was developed. In early measurements~\citep{19533Powell}, it was found that the screech frequency experienced abrupt changes as the inlet pressure increased.
This frequency jumping phenomenon is commonly referred to as mode staging, and four different stages, i.e. stages A, B, C, and D were observed by Powell, among which the stage A can be further divided into two stages named $\rm{A}_{1}$ and $\rm{A}_{2}$~\citep{M.Merle}. 
\add{It was found that the azimuthal mode of both sound and instability waves changes as the mode staging occurs,} which shows a strong sensibility to the facility and initial conditions~\citep{1969V.M.ANUFRIEV, E.GUTMARK1990, Panda_standingwave_2}, and the switch from one mode to another is nearly immediate~\citep{Nagel_modestaging}. Despite different stages show different characteristics, it is interesting to note that they can appear simultaneously in one jet flow~\citep{1996AIAA_Raman}. 
To interpret mode staging, \citet{Tam_and_shen} suggested that it was the neutral acoustic waves, rather than free acoustic waves, that complete the feedback loop in $\rm{A}_{2}$ 
and B modes. \add{Recent works~\citep{Gojon_modeStaging,edgington_modestaging,Xiang-Ru_2020,NewC_1} showed that both the $\rm{A}_{1}$ and $\rm{A}_{2}$ modes were closed by the
neutral acoustic waves~(or guided-jet modes).} 
\citet{X.D.Li} used the original phase condition shown in~(\ref{phase_condition}) and inferred the value of $N$ in their numerical study. They showed that $N$ differed across various stages, and when  this difference was considered the prediction  was in very good agreement with the experimental data. However, the mechanism behind this mode transition and its high sensitivity to the initial conditions are yet to be clarified. \add{In addition, it was found that nonlinearity could arise during the mode staging in circular jets~\citep{nonlinear_in_staging} and  screech tones were not independent but were instead nonlinearly phased locked to each other in rectangular jets~\citep{1997_POF_SHWalker}, which increases difficulties for its modeling.}

In addition to the screech frequency and mode staging, noise directivity is the second characteristic that has been widely studied. 
It was found that the
acoustic radiation at the fundamental frequency appeared strongest in the upstream
direction, whereas at the harmonic frequency there was a strong beaming to the
side of the jet~\citep{19533Powell}. To explain this, Powell proposed the monopole array theory,
which was generally in good agreement with experiment results. The
directivity of the fundamental tone and its harmonics in supersonic round nozzles were measured by~\citet{1983Norm}, the results of which were compared with the monopole
array theory when nine monopoles with a parabolic intensity
distribution were considered. One particularly interesting observation was that the strongest emission appeared somewhere near $150^\circ$, not $180^\circ$ to the downstream jet axis. A quick decay
occurred when the observer angle approached $180^\circ$, which could not be
predicted by Powell's model. As a matter of fact, if all the sound sources were
monopoles and the frequency of the fundamental tone was obtained by~(\ref{equ_screech f}), the acoustic radiation would be the strongest at $180^\circ$.
Following Norum's idea, the directivity pattern of three equal-spaced monopoles of various intensities was studied by~\citet{M.Kandula}. It seemed that this variation did not affect the location of the main directivity lobe. The temperature influence on the directivity pattern~\citep{heateffect} was also investigated in round jets, but no significant difference was found from the unheated one. In the case of rectangular nozzles, the screech problem may become more complicated compared to axisymmetric round jets. However, this problem can be greatly simplified when the rectangular nozzle is of high aspect ratios, in which case it reduces to a two-dimensional problem. Numerical simulations~\citep{numerical_directivity_of_rectangular} and experiments~\citep{1997_POF_SHWalker} were conducted to study the directivity patterns of the screech and its harmonics in rectangular jets of high aspect ratios and the results were similar to that of~\citet{19533Powell}. Recently, \citet{2014TAM} considered another nonlinear interaction
mechanism between shock, instability, and acoustic waves. They then proposed a model to predict the lobe
position in the directivity patterns.
The result was in good agreement with the experimental data~\citep{1983Norm} at harmonic frequencies, but appeared less so for the fundamental tone, in particular for the lobe position.

As argued by Powell, four stages were involved in the screech cycle. Among the four stages, it is believed that the interaction between the shock and instability waves plays a critical role in understanding the physics \adddd{of screech}. Not only because this interaction produces sound that is directly measurable, but also because it is the key to understand the noise generation mechanism. Despite its importance, not many theoretical models are proposed to predict the interaction. The reason is in part due to the complex flow nature present in the interaction~\citep{TAmaning}, especially when the shock waves are intense in highly underexpanded and overexpanded jet flows. ~\citet{1973Harper} used Powell's phased-array model to study the interaction between the disturbance in jet shear
layers and the shock cells. The frequency of the emitting sound was obtained. Subsequently,~\citet{1988Tam} developed a shock cell model composed 
of time-independent waveguide modes~\citep{1985Tam}. The turbulence structures were modeled using a  noise initiated at the shear layer at the nozzle lip. The weak interaction between these two components then gave rise to the sound field. \add{Although the sound generated by shock-vortex interaction was analytically studied in the first part of the work, it was remarked by the author that~\citep{1988Tam} “enormous amounts of numerical computations are
required" for practical calculations. Thus, “a model source function" was used instead in light of the extreme complexity of the practical evaluation of the formulation to calculate the directivity patterns of BBSAN}.~\citet{2005Lele} further developed Tam's theory. Based on the method proposed by~\citet{1952lighthill}, he used the wavepacket model to describe the instability waves initialized by the white noise, and a vortex sheet model was utilized to calculate the shock cell structure. These two components were inserted into the Helmholz equation as the source term. The sound field was obtained by integration. This model was used in subsequent numerical simulations~\citep{2019Wong}. However, the sound sources in the above-mentioned models were obtained by a simple combination of the shock and instability waves. A correct source term directly from governing equations would be more desirable.~\add{A somewhat different approach to model the sound generation is the so-called shock leakage mechanism. It was proposed by~\citet{SKLele_TAManning_shockleakage, TAmaning}, and theoretically developed by~\citet{2003Suzeki} and~\citet{KS_TAManning_shockleakage}. Recently, it was experimentally 
observed by~\citet{edgington_shockleakage}. In addition, a very recent work~\citep{absolute_instability} numerically studied the linear stability characteristics of shock-containing jet flows, where the shock was assumed to be of small amplitude and a sinusoidal form. It was found that the characteristics of the instability waves in shock-containing jets are different from those in shock-free supersonic jets, and a new interpretation of screech was proposed based on this observation.}

In 1994,~\citet{1995Kerchen} developed \adddd{an analytical} shock
instability-wave interaction model for 2D planar vortex sheet flows. The source term was obtained directly from the governing equations. One shock cell was considered to interact with the instability waves near the vortex sheet. However, it
was found that the radiation field peaked at $48^\circ$ to the downstream jet axis, which contradicted experimental observations. Despite of numerous attempts, an analytical and quantitative study of the interaction
between the shock and instability waves, which is capable of predicting not only screech
frequencies, but also directivity patterns of screech tones,
appears yet to be seen. This paper aims to
develop such a model to predict the sound arising from the interaction
between shock and instability waves. The model follows the
asymptotic expansion method proposed by~\citet{1995Kerchen}, but a more realistic jet and shock cell structures are considered. 
This paper is structured as follows. Section~\ref{sec:types_paper} presents a detailed analytical derivation of the model, while section~\ref{section:results and discussion} shows the prediction of the screech frequency and the directivity patterns of the fundamental tone and its
harmonics. The near-field pressure and noise generation mechanism are subsequently discussed.  
Conclusions are presented in section~\ref{section:conclusion}.

\section{Analytical formulation}
\label{sec:types_paper}
\subsection{The interaction model}
\label{subsection:base assumptions}
\begin{figure}
   \centering
   \includegraphics[width = 0.65\textwidth]{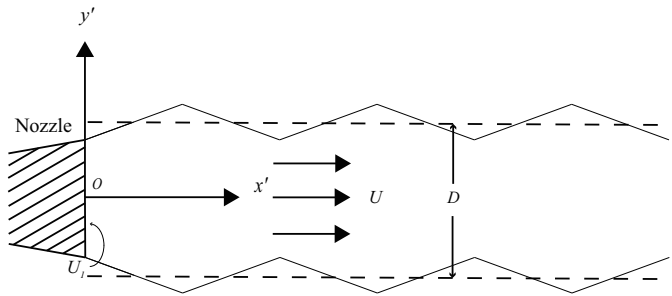}
   \caption{The schematic of the vortex-sheet flow configuration and Cartesian
   coordinates. The origin is fixed at the centre of the nozzle while $x^\prime$ and
   $y^\prime$ represent the streamwise and cross-flow coordinates, respectively.}
\label{fig:example1}
\end{figure}

To enable analytical progression, we start with a vortex sheet model. As shown
in figure~\ref{fig:example1}, the coordinate axes $(x^\prime,y^\prime)$ are
chosen to be parallel and perpendicular to the nozzle centreline, respectively. Here $D$ is the jet height~(note that $D$ is generally not equal to the height of the nozzle, and we take the height of the fully-expanded jet as the base flow height~\citep{tam_diameter}). $U_{1}$ is the jet velocity at the nozzle exit plane, while $U$ is
the velocity of the fully-expanded jet after exiting from the nozzle. The fully-expanded base flow described in figure~\ref{fig:example1} takes the form
\begin{equation}
  \boldsymbol{u}_{0}= 
    \begin{cases}
      0, & |y^\prime|>D/2 \\
      U \mathbf{e}_{x^\prime}, &  |y^\prime|\leq D/2,
  \end{cases} 
\end{equation}
where $\mathbf{e}_{x^\prime}$ is the unit vector in the $x^\prime$ direction. We assume that the shock and
instability waves are  of small amplitudes and can be linearized around the base flow and described by
linear theories. Of course, it is known that the interaction between the shock
and instability waves primarily occurs several shock cells downstream from the
nozzle exit~\citep{1994Suda,S.kaji,Sources}. At these locations, instability waves are
likely to grow to a significant amplitude where nonlinear effects become
important and the instability waves may start to saturate and even decay.
However, it is known that linear theories can predict the wavelength of these
large coherent structures well beyond the linear stage~\citep{Crow_Champion, 2013_annual_rev,2021Edington}. We may, therefore, use linear stability analysis to determine the
wavelength (hence convection velocity) of instability waves, which are
particularly important for the generation of screech tones. The linear growth
rate calculated describes the early evolving of the instability wave and its role is
discussed separately in subsequent modelling. The interaction between the shock and instability waves several shock-cells downstream are likely to be nonlinear in strongly underexpanded jets. However, a linear interaction model may suffice to describe the interaction between weak shock and instability waves. Besides, similar to the successful prediction of the wavelength and convection velocities of large coherent structures by linear theories, a linear theory may still possess the many essential features of a nonlinear jet screech. We therefore start with a linear interaction between the shock and instability waves. With these
assumptions we start to seek an analytical model describing noise generation
due to the interaction between shock and instability waves.

Given the fact that the Reynolds number is high and the instability waves are
essentially inviscid, we start from the Euler equations shown as follows
\begin{equation}
    \frac{\mathrm{D} \rho}{\mathrm{D} t^\prime } + \add{\rho}\bnabla \cdot \boldsymbol{u}=0,
\end{equation}
\begin{equation}
    \rho \frac{\mathrm{D}\boldsymbol{u}}{\mathrm{D} t^\prime}= -\bnabla p,
\end{equation}
\begin{equation}
    \frac{\mathrm{D} s}{\mathrm{D} t^\prime} = 0,
\end{equation}
where $t^{\prime}$ denotes time, $\boldsymbol{u}=(u,v)$ the velocity, $p$ the
pressure, $\rho$ the density, $s$ the entropy, and \adddd{$\mathrm{D}/\mathrm{D} t^\prime=\partial/\partial t^\prime+ \boldsymbol{u}\cdot\bnabla$}.   
Because the entropy increase
across a weak shock is a high-order small term~\citep{1995Kerchen}, the
isentropic condition is used here. 

To determine the solution, both kinematic and dynamic boundary conditions need to be satisfied across the vortex sheet. The dynamic boundary conditions read
\begin{equation}
    \label{equP}
\left.p_{+}\right|_{y^\prime = h^{'}} = \left.p_{-}\right|_{y^\prime = h^\prime},
\end{equation}
while the kinematic boundary condition requires
\begin{equation}\label{equB1}
\left. v_{+}\right|_{y^\prime = h^\prime} =\left(\frac{\partial h^\prime}{\partial t^\prime}+\left. u_{+}\right|_{y^\prime = h^\prime}\frac{\partial h^\prime}{\partial x^\prime}\right),
\end {equation}
\begin{equation}\label{equB2}\left.
v_{-}\right|_{y^\prime = h^\prime} =\left(\frac{\partial h^\prime}{\partial t^\prime}+\left. u_{-}\right|_{y^\prime = h^\prime}\frac{\partial h^\prime}{\partial x^\prime}\right),
\end{equation}
where $(\cdot)_+$ and $(\cdot)_-$ represent the quantities outside and within the jet flow respectively, and $h^\prime$ denotes the displaced $y^\prime$ of the vortex sheet.

Following \citet{1995Kerchen}, we use $\delta$ and $\epsilon$ to denote the strength of the shock and instability waves, respectively. These two parameters are assumed to be of the small magnitude representing small perturbation compared to the mean jet flow. \add{Note that although shock waves can be intense in strongly imperfectly-expanded jets, they could also be infinitesimally weak when the wave angle \adddd{approaches the Mach angle}~\citep{Anderson}. In this model, we consider a slightly imperfectly expanded jet, in which case the shock-associated perturbations can be relatively weak, thus a linear model may be developed~\citep{tam_machwave,1985Tam,  1995Kerchen}}. The velocity field can be expanded
using these two parameters as \citep{1995Kerchen} 
\begin{equation}
   \boldsymbol{u}=\boldsymbol{u}_{0}
   +\delta \boldsymbol{u}_{m}
   +\epsilon\boldsymbol{u}_{v}
   +\delta^{2}\boldsymbol{u}_{m2}
   +\epsilon^{2}\boldsymbol{u}_{v2}
   +\delta\epsilon\boldsymbol{u}_{i}+...\ ,
\end{equation}
where $\boldsymbol{u}_{0}$ is the mean \adddd{velocity}, $\boldsymbol{u}_{m}$
represents the linear perturbation due to shock waves, and $\boldsymbol{u}_{v}$
is the linear unsteady perturbation due to instability waves. For higher
orders, the $\delta^2$ term represents nonlinear steady modification of the
shock waves and is independent of time. As we are interested in sound
generation, this term can be neglected. The second-order term
$\epsilon^2$ represents the nonlinear correction to the linear stability waves, and
we neglect it by only considering the leading-order contribution from the
$\epsilon$ term. The omission of this term follows immediately if $1 \gg \delta
\gg \epsilon$ is assumed, which implies $\epsilon^2$ is the highest-order term
of the expansion. The $\delta \epsilon$ term represents the interaction due to
shock and stability waves, as a result of which sound is generated. \add{Physically, this entails that the shock and instability waves that interact to produce sound in realistic flows may be approximated by the linear shock and instability solutions to the base flow.} The
pressure, density, and vortex sheet displacement have similar expansions, i.e.
\begin{equation}
   p=p_{0}+ \delta p_{m} +\epsilon p_{v}+\delta^{2}p_{m2}+\epsilon^{2}p_{v2}+\delta\epsilon p_{i}+...\ ,
\end{equation}
\begin{equation}
   \rho=\rho_{0}+ \delta\rho_{m} +\epsilon\rho_{v}+\delta^{2}\rho_{m2}+\epsilon^{2}\rho_{v2}+\delta\epsilon\rho_{i}+...\ ,
\end{equation}
\begin{equation}
   h^\prime=h^\prime_{0}+ \delta h^\prime_{m} +\epsilon h^\prime_{v}+\delta^{2}h^\prime_{m2}+\epsilon^{2}h^\prime_{v2}+\delta\epsilon h^\prime_{i}+...\ .
\end{equation}

The \adddd{mean} temperature across the jet flow is
different, while the \adddd{mean} pressure $p_{0}$ \adddd{remains identical} in both regions. If a perfect gas is assumed, then $c_0=\sqrt{\gamma p_0/\rho_0}$ can be used to calculate the \adddd{mean} speed of
sound inside and outside the jet, and it is straightforward to show that
$\rho_{0-}c_{0-}^{2}=\rho_{0+}c_{0+}^{2}$. We define $M_{-}=U/c_{0-}$ and
$M_{+}=U/c_{0+}$ to denote the \adddd{mean} Mach numbers inside and outside the jet,
respectively. Substituting all the expansions to the Euler equation and
boundary conditions, and collecting the terms $O(\delta)$, $O(\epsilon)$ and
$O(\delta\epsilon)$, we obtain the equations governing the shock, instability waves and
their interaction, respectively.

\subsection{The shock model}
The $O(\delta)$ terms in the governing equations representing the linear perturbation induced by the shock wave satisfy
\begin{equation}
    \frac{\mathrm{D}_{0}p_{m}}{\mathrm{D} t^{'}}+\rho_{0}c_{0}^{2}\bnabla\cdot\boldsymbol{u}_{m}=0,
\end{equation}
\begin{equation}
    \rho_{0}\frac{\mathrm{D}_{0}\boldsymbol{u}_{m}}{\mathrm{D} t^{'}}=-\bnabla p_{m},
    \label{equ_shock_momentum}
\end{equation}
where \adddd{$\mathrm{D_0}/\mathrm{D} t^\prime$ denotes $\partial/\partial t^\prime+ \boldsymbol{u}_0\cdot\bnabla$}. These two equations can be combined to yield
\begin{equation}
    \bnabla^{2}p_{m}-M^{2}\frac{\partial^{2}p_{m}}{\partial x^{\prime2}}=0.
    \label{equ:waveequation}
\end{equation}
\adddd{Equation (\ref{equ:waveequation}) reduces to the Laplace equation outside the jet, since the mean flow velocity is zero there. Within the jet, by defining $\beta=\sqrt{M_{-}^{2}-1}$, we can rewrite~(\ref{equ:waveequation})
to be}
\begin{equation}
    \frac{\partial^{2}p_{m-}}{\partial y^{\prime2}}-\beta^{2}\frac{\partial^{2}p_{m-}}{\partial x^{\prime2}}=0.
    \label{equ:waveequation2}
\end{equation}

We see that the linear pressure field \adddd{within the jet} induced by weak shock waves satisfies the
wave equation. Such an equation can admit many solutions subject to different
boundary conditions; for illustration purposes, \citet{1995Kerchen} used a step
function for his single planar vortex sheet. In order to have a much more realistic shock cell structure, we use
Pack's model~\citep{1950Pack} in this paper. Note that we use the jet height $D$ here, instead of the nozzle height, to nondimensionalize the streamwise and cross-flow coordinates, i.e. $x=x^\prime/D$, $y=y^\prime/D$. The velocity potential perturbation \adddd{within the jet} induced by the shock waves can be
written as~\citep{1950Pack, tam_diameter}
\begin{equation}
    \phi=\sum_{j=1}A_{j}\cos(\beta a_{j} y)\sin(a_{j} x),
    \label{Pack's model}
\end{equation}
where $j$ represents the $j$th mode of the shock wave. Two parameters $A_{j}$
and $a_{j}$ in the equation above \adddd{are}
\begin{equation}
    A_{j}=(-1)^{j}\frac{4\beta}{\pi^{2}}\frac{\mathcal{U}}{(2j-1)^2},
    \label{equ:A}
\end{equation}
\begin{equation}
    a_{j}=\frac{(2j-1)\pi}{\beta}.
\end{equation}
The constant $\mathcal{U}$ in the above equation is defined by  \add{$\mathcal{U}=U_{1}-U.$} We see from~(\ref{equ:A}) that the amplitude of this potential function decreases quickly as the
mode number increases. In light of the linearity of the model, it is convenient
to consider each mode separately. In the present study, we focus on the
leading-order mode. Higher-order terms can be easily included at a later stage
should necessity \add{arise}. For the leading-order mode, the shock wave is
periodically distributed along the streamwise direction with a shock spacing \adddd{$s=2\pi/a_{1}.$} In what follows, the subscripts in parameters $A_{1}$ and $a_{1}$ are omitted for
clarity.

\add{The corresponding velocity, pressure, and the vortex sheet deflection at the boundary of the jet flow are shown in Appendix~\ref{appA}.}
\subsection{The instability waves} 
\label{sec:filetypes} 
Similar to the derivation of the shock equations, the governing equations for
the instability waves can be obtained by collecting the $O(\epsilon)$ terms,
i.e.
\begin{equation}
    \frac{\mathrm{D}_{0}p_{v}}{\mathrm{D} t^\prime}+\rho_{0}c_{0}^{2}\bnabla\cdot\boldsymbol{u}_{v}=0,
    \label{equ_v_continuty}
\end{equation}
\begin{equation}
    \rho_{0}\frac{\mathrm{D}_{0}\boldsymbol{u}_{v}}{\mathrm{D} t^\prime}=-\bnabla p_{v}.
    \label{equ_v_momentum}
\end{equation}
Considering the $O(\epsilon)$ terms in the boundary conditions shown in
(\ref{equP}), (\ref{equB1}) and (\ref{equB2}), we see that the two matching
conditions can be linearised to the dynamic and kinematic conditions on $y=\pm
D/2$, i.e.
\begin{equation}
    p_{v+}=p_{v-}, 
\end{equation}
\begin{equation}
        v_{v+}=\frac{\partial h^\prime_{v}}{\partial t^\prime},\quad v_{v-}=\frac{\partial h^\prime_{v}}{\partial t^\prime}+U\frac{\partial h^\prime_{v}}{\partial x^\prime},
        \label{equ_v_continuity of displacement}
\end{equation}
where $h^\prime_{v}$ denotes the disturbed height of the vortex sheets due to the
instability waves. Since the initial base flow is irrotational and
inviscid both inside and outside of the vortex sheet, the linear perturbation
can be expressed as a velocity potential $\phi_{v}$. The continuity
equation~(\ref{equ_v_continuty}), and the momentum equation~(\ref{equ_v_momentum}), can
be combined to yield 
\begin{equation}
    \bnabla^{2}\phi_{v}-\frac{1}{c_{0}^{2}}\frac{\mathrm{D}_{0}^{2}\phi_{v}}{\mathrm{D} t^{\prime2}}=0.
\end{equation}
Similarly the dynamic boundary condition reduces to
\begin{equation}
\rho_{0+}\frac{\partial \phi_{v+}}{\partial t^\prime}=\rho_{0-}\left[\frac{\partial \phi_{v-}}{\partial t^\prime}+U\frac{\partial \phi_{v-}}{\partial x^\prime} \right].
\end{equation}
For the kinematic boundary condition, the two equations shown in
(\ref{equ_v_continuity of displacement}) can be combined to yield
\begin{equation}
    \left(\frac{\partial}{\partial t^\prime}+U\frac{\partial}{\partial x^\prime}\right)\frac{\partial \phi_{v+}}{\partial y^\prime}=\frac{\partial^{2}\phi_{v-}}{\partial t^{'} \partial y^\prime}.
\end{equation}

With temporal and spatial harmonic assumptions, the perturbations induced by the
instability waves have the form of
   \add{$ \mathrm{e}^{{\mathrm i}(\alpha^\prime x^\prime-\omega^\prime t^\prime)},$} where the spatial wavenumber $\alpha^\prime$ is complex and the eigenvalue with a 
negative imaginary part represents instability. We use the velocity of the fully-expanded jet flow $U$ to nondimensionalize
other variables, for instance the nondimensional time, frequency and wavenumber
are \add{$t=t^\prime U/D, \omega=\omega^\prime D/U, \alpha=D\alpha^\prime$}\adddd{, respectively.} 


Combining the governing equations and boundary conditions, and noticing that the base flow outside the jet is 0, we find that the velocity potential can be expressed as
\begin{equation}
    \phi=U D \mathrm{e}^{i(\alpha x-\omega t)}\times
    \begin{cases}
      \frac{1}{M_{+}^{2}}\mathrm{e}^{-m_{+}y}, & y>\frac{1}{2} \\[2pt]
      \frac{1}{M_{-}^{2}}\frac{\omega}{\omega-\alpha}(k_{1}\mathrm{e}^{m_{-}y}+k_{2}\mathrm{e}^{-m_{-}y}),& y\leq |\frac{1}{2}|\\
     \frac{1}{M_{+}^{2}}k_{3}\mathrm{e}^{m_{+}y},& y<-\frac{1}{2},
    \end{cases} 
    \label{equ:instabilty_potential}
\end{equation}
where $ m_{+}=\sqrt{\alpha^{2}-\omega^{2}M_{+}^{2}},\
m_{-}=\sqrt{\alpha^{2}-M_{-}^{2}(\omega-\alpha)^{2}}$. The branch cut is chosen such that
the real part of $m_{+}$ is positive. $k_{1}, k_{2}, k_{3}$ are undetermined
coefficients. It can be seen that both antisymmetric and symmetric modes can
exist, corresponding to $k_{3}=- 1$ and $k_{3}=1$, respectively. Using the dynamic and kinematic boundary conditions, the dispersion
relations can be found to be
\begin{equation}
    \mathrm{e}^{2m_{-}}=\frac{(\omega^{2}m_{-}/M_{-}^{2}-(\omega-\alpha)^2m_{+}/M_{+}^{2})^{2}}{(\omega^{2}m_{-}/M_{-}^{2}+(\omega-\alpha)^2m_{+}/M_{+}^{2})^{2}}.
    \label{equ:dispersion relationship}
\end{equation}
For the symmetric mode, this reduces to
\begin{equation}
    {\rm tanh}(\frac{m_{-}}{2})\frac{\omega^{2} m_{-}}{M_{-}^{2}(\omega-\alpha)^{2}}+\frac{m_{+}}{M_{+}^{2}}=0,
    \label{dispersion_sym}
\end{equation}
where the parameters attain the following values
\begin{equation}
    k_{1}=k_{2}=\frac{\mathrm{e}^{-\frac{1}{2}m_{+}}}{2{\rm cosh}(\frac{1}{2}m_{-})}, \quad k_{3}=1.
\end{equation}
For the antisymmetric mode, the dispersion relationship reduces to
\begin{equation}
    \frac{\omega^{2} m_{-}}{M_{-}^{2}(\omega-\alpha)^{2}}+{\rm tanh}(\frac{m_{-}}{2})\frac{m_{+}}{M_{+}^{2}}=0,
    \label{dispersion_anti}
\end{equation}
where the three parameters take the values of
\begin{equation}
    k_{1}=-k_{2}=\frac{\mathrm{e}^{-\frac{1}{2}m_{+}}}{2{\rm sinh}(\frac{1}{2}m_{-})},\quad k_{3}=-1.
\end{equation}
\add{ The corresponding pressure, velocity, and the deflection of the jet boundary due to the instability waves are shown in Appendix~\ref{appB}}. We can see that the deflection generated by the instability wave at the upper and lower boundaries are symmetric and antisymmetric for the symmetric and antisymmetric modes, respectively, as would be expected. 

Experiments found that rectangular jets are capable of sustaining both symmetric
and antisymmetric  \adddd{oscillation} modes~\citep{1994Suda,S.kaji}, which can be directly linked to the instability of the jet. In our analysis, both symmetric and antisymmetric instability modes can be considered. \adddd{But for jet flows from high-aspect-ratio rectangular nozzles, the flapping mode is dominant~\citep{2019EDINGTON} and the problem can be approximated by a 2D theory}. So in what follows only the antisymmetric mode of instability waves is considered.

\subsection{The interaction between shock and instability waves} 
\label{subsec:sound}
Having obtained the shock and instability waves, we are now in a position to
consider the $O(\delta\epsilon)$ terms in the governing equations. When a
perfect gas is assumed \adddd{($\rho_0 c_0^{2}=\gamma p_0$)}, the continuity and momentum
equations can be expressed as
\begin{equation}
        \frac{\mathrm{D}_{0}p_{i}}{\mathrm{D} t^\prime}+\rho_{0}c_{0}^{2}\bnabla\cdot\boldsymbol{u}_{i}=-[\boldsymbol{u}_m\cdot\bnabla p_{v}+\boldsymbol{u}_{v}\cdot\bnabla p_{m}+\gamma p_{m}\bnabla\cdot\boldsymbol{u}_v+\gamma p_{v}\bnabla \cdot \boldsymbol{u}_{m}],
        \label{equ_sound_continuity}
\end{equation}
\begin{equation}
        \rho_{0}\frac{\mathrm{D}_{0}\boldsymbol{u}_{i}}{\mathrm{D} t^\prime}+\bnabla p_{i}=-[\rho_{0}(\boldsymbol{u}_m\cdot\bnabla \boldsymbol{u}_{v}+\boldsymbol{u}_{v}\cdot\bnabla \boldsymbol{u}_{m})+\rho_{m}\frac{\mathrm{D}_{0}\boldsymbol{u}_v}{D t^\prime}+\rho_{v}\frac{\mathrm{D}_{0}\boldsymbol{u}_m}{D t^\prime}].
        \label{equ_sound_momentum}
\end{equation}
Substituting the momentum equations for the shock wave, i.e.
(\ref{equ_shock_momentum}), and the instability wave, i.e. (\ref{equ_v_momentum}),
into~(\ref{equ_sound_momentum}), we have
 \begin{equation}
      \rho_{0}\frac{\mathrm{D}_{0}\boldsymbol{u}_{i}}{\mathrm{D} t^\prime}+\bnabla p_{i}=-[\rho_{0}\bnabla(\boldsymbol{u}_{m}\cdot\boldsymbol{u}_{v})+\frac{1}{\rho_{0}c_{0}^{2}}\bnabla(p_{m}p_{v})].
      \label{18}
  \end{equation}
   The interaction field is also irrotational, i.e. $u_{i}=\bnabla \phi_{i}$.
   Then~(\ref{18}) can be integrated to obtain
   \begin{equation}
      p_{i}= -\rho_{0}\frac{\mathrm{D}_{0}\phi_{i}}{\mathrm{D} t^\prime}-\rho_{0}\boldsymbol{u}_{m}\cdot\boldsymbol{u}_{v}+\frac{1}{\rho_{0}c_{0}^{2}}p_{m}p_{v}.
      \label{20}
  \end{equation}
  Combining~(\ref{equ_sound_continuity}) and (\ref{20}), and considering the momentum and
  continuity equations for the $O(\alpha)$ and $O(\epsilon)$ terms, we find
  that $\phi_{i}$ satisfies the following inhomogeneous wave equation
  \begin{equation}
      \bnabla^{2}\phi_{i}-\frac{1}{c_{0}^{2}}\frac{\mathrm{D}_{0}^{2}\phi_{i}}{\mathrm{D} t^{\prime2}}=\frac{-1}{\rho_{0}c_{0}^{2}}[2(\mathbf{u}_m\cdot\bnabla p_{v}+\boldsymbol{u}_v\cdot\bnabla p_{m})+(\gamma-1)(p_{m}\bnabla\cdot\boldsymbol{u}_{v}+p_{v}\bnabla\cdot\boldsymbol{u}_{m})].
      \label{equ_sound_inhomogenous}
        \end{equation}
        
  Next we consider the continuity of the pressure across
  the vortex sheet. Similar to \citet{1995Kerchen}, on the two boundaries
  $y^\prime=1/2D$ and $y^\prime=-1/2D$, the dynamic boundary
  condition reduces to
   \begin{equation}
      p_{i+}+h_{m}^\prime\frac{\partial p_{v+}}{\partial y^\prime}= p_{i-}+h_{m}^\prime\frac{\partial p_{v-}}{\partial y^\prime}+h_{v}^\prime\frac{\partial p_{m-}}{\partial y^\prime},
  \end{equation}
  while the kinematic boundary condition requires
  \begin{equation}
         v_{i+}+\frac{\partial v_{v+}}{\partial y^\prime}h^\prime_{m}=\frac{\partial h_{i}^\prime}{\partial t^\prime}+u_{v+}\frac{\partial h_{m}^\prime}{\partial x^\prime},
  \end{equation}
  \begin{equation}
      v_{i-}+\frac{\partial v_{v-}}{\partial y^\prime}h^\prime_{m}+\frac{\partial v_{m-}}{\partial y^\prime}h^\prime_{v}=\frac{\partial h_{i}^\prime}{\partial t^\prime}+U\frac{\partial h^\prime_{i}}{\partial x^\prime}+u_{v-}\frac{\partial h_{m}^\prime}{\partial x^\prime}+u_{m-}\frac{\partial h_{v}^\prime}{\partial x^\prime}.
  \end{equation}
  Note that outside the jet flow, there is no
  perturbation due to shock waves, therefore the right-hand side of
  (\ref{equ_sound_inhomogenous}) vanishes. This equation degenerates to a homogeneous
  wave equation. Using the same $U$ and $D$ to
  nondimensionalize the velocity potential, we obtain
  \begin{equation}
      \phi_{i+}=2 U D g_{+}\mathrm{e}^{-\mathrm{i}\omega t},
      \label{potential_outside}
  \end{equation}
  \begin{equation}
           \phi_{i-}=2 U D g_{-}\mathrm{e}^{-\mathrm{i}\omega t},
  \end{equation}
  \add{where $g_\pm$ are the nondimensionalized potential functions.} 
  Outside the jet flow, the base flow is uniformly 0, and (\ref{equ_sound_inhomogenous}) reduces to
    \begin{equation}
      \frac{\partial^{2}g_{+}}{\partial x^{2}}+\frac{\partial^{2}g_{+}}{\partial y^{2}}+\omega^{2}M_{+}^{2}g_{+}=0.
  \end{equation}
  Inside the jet flow, $g_{-}$ satisfies
    \begin{eqnarray}
    \frac{\partial^{2}g_{-}}{\partial x^{2}}+\frac{\partial^{2}g_{-}}{\partial y^{2}}+M_{-}^{2}(\omega+\mathrm{i}\frac{\partial}{\partial x})^{2}g_{-} & = &2k_{1}\mathrm{e}^{\mathrm{i} \alpha x}\frac{A}{U}\bigg[ 
    \sinh(\zeta_{1}y)(B_{1}\cos(a x)+B_{2}\sin(a x))\nonumber\\
    & + &\sinh(\zeta_{2}y)(B_{3}\cos(a x)+B_{4}\sin(a x)\bigg],
    \label{equ_sound_inhomogenous2}
  \end{eqnarray}
where
\begin{gather}
    \zeta_{1}=m_{-}+\mathrm{i} a\beta,\nonumber\\
    \zeta_{2}=m_{-}- \mathrm{i}a\beta,
\end{gather}
\begin{gather}
    B_{1}=\frac{a\omega}{2}\left(\alpha+\mathrm{i}\frac{a\beta m_{-}}{\omega-\alpha}-\frac{\gamma-1}{2}(\omega-\alpha)M_{-}^{2} \right),\nonumber\\
    B_{2}=\frac{a\omega}{2}\left(m_{-}\beta-\mathrm{i}\frac{\alpha a}{\omega-\alpha}+\frac{\gamma-1}{2}\mathrm{i}a M_{-}^{2} \right),\nonumber\\
    B_{3}=\frac{a\omega}{2}\left(\alpha-\mathrm{i}\frac{a\beta m_{-}}{\omega-\alpha}-\frac{\gamma-1}{2}(\omega-\alpha)M_{-}^{2} \right),\nonumber\\
    B_{4}=\frac{a\omega}{2}\left(-m_{-}\beta-\mathrm{i}\frac{\alpha a}{\omega-\alpha}+\frac{\gamma-1}{2}\mathrm{i}a M_{-}^{2} \right).
\end{gather}
Besides, the two boundary conditions can be reorganized as
\begin{eqnarray}
        \omega g_{+}-\frac{M_{-}^{2}}{M_{+}^{2}}(\omega+\mathrm{i}\frac{\partial}{\partial x})g_{-}&=&\pm\frac{A\beta}{2UM_{+}^{2}}\sin(\frac{1}{2}a\beta )\cos(a x)\bigg[\omega\big(2k_{1}m_{-}\cosh(\frac{1}{2}m_{-})\nonumber\\
        &\pm& m_{+}\mathrm{e}^{\mp\frac{1}{2}m_{+}}\big)
        -\frac{\mathrm{e}^{-\frac{1}{2}m_{+}}}{\omega}m_{+}\frac{M_{-}^{2}}{M_{+}^{2}}a^{2}\bigg]\mathrm{e}^{\mathrm{i} \alpha x}
        \label{equ:boundary condition 1}
\end{eqnarray}
and
\begin{eqnarray}
    (1+\frac{\mathrm{i}}{\omega}\frac{\partial}{\partial x})\frac{\partial g_{+}}{\partial y}-\frac{\partial g_{-}}{\partial y}&=&\frac{ A }{2 U}\bigg[{\mathrm{i}}a\beta\sin(a x)\big[b_{1}\sin(\pm\frac{1}{2}a\beta )+c\beta\cos(\frac{1}{2}a\beta ) \big] \nonumber\\
    &+&\cos(a x)\big[ b_{2}\sin(\pm\frac{1}{2}a\beta )-c \alpha\cos(\frac{1}{2}a\beta ) \big]\bigg]\mathrm{e}^{\mathrm{i} \alpha x},
    \label{equ:boundary condition 2}
\end{eqnarray}
where the upper and lower of signs of $\pm$ and $\mp$ correspond to the matching conditions on $y=1/2$ and $y=-1/2$, respectively, and
\begin{gather}
    b_{1}=\pm\frac{2\alpha k_{1}\sinh(\frac{1}{2}m_{-} )}{M_{+}^{2}\omega}[(\alpha-\omega)+\frac{m_{+}^{2}}{\alpha}]\pm\frac{\alpha}{M_{-}^{2}}\frac{\omega}{\omega-\alpha}\mathrm{e}^{\mp\frac{1}{2}m_{+}},\\
        b_{2}=\pm\frac{2\beta k_{1}\sinh(\frac{1}{2}m_{-} )}{M_{+}^{2}\omega}[(\alpha-\omega)m_{+}^{2}+\alpha a^{2}]\pm m_{-}^{2}\frac{\beta}{M_{-}^{2}}\frac{\omega}{\omega-\alpha}\mathrm{e}^{\mp\frac{1}{2}m_{+}},\\
        c=\frac{a}{\omega}\frac{m_{+}}{M_{+}^{2}}\mathrm{e}^{-\frac{1}{2}m_{+}}.
\end{gather}

To obtain the solution to these equations, Fourier \add{transform} is used. The Fourier transforms $G_{\pm}(\lambda,y)$ are defined as
    \begin{equation}
	G_{\pm}(\lambda,y)=\int_{-\infty}^{+\infty} g_{\pm}(x,y)\mathrm{e}^{\mathrm{i} \lambda x}\mathrm{d} x,
	\label{equ:fourier}
    \end{equation}
\add{where $\lambda$ is the wavenumber in the streamwise direction.} Outside the jet flow, it is easy to find that $G$
satisfies
    \begin{equation}
        G_{+}(\lambda,y)=
    \begin{cases}    
        D_{1}(\lambda)\mathrm{e}^{-\gamma_{+}y}, &\test{y>1/2}\\[2pt]
        D_{4}(\lambda)\mathrm{e}^{\gamma_{+}y}, &\test{y<-1/2},
        \end{cases}
    \end{equation}
where
\begin{equation}
    \gamma_{+}(\lambda)=\sqrt{\lambda^{2}-\omega^{2}M_{+}^{2}},
\end{equation}
and $D_{1}$ and $D_{4}$ are two undetermined coefficients related to $\lambda$.
    	\begin{figure}
		\centering
		\includegraphics[width = 0.4\textwidth]{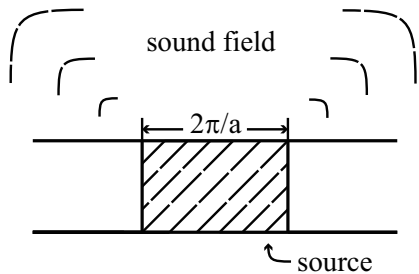}
		\caption{The schematic of an effective source term located within one single
		shock cell. The total sound field is
		equivalent to a linear superimposition of the results from a number of
		shock cells.}\label{fig:schematic}
	\end{figure}

When Fourier \add{transform} is used to the source term of
(\ref{equ_sound_inhomogenous2}),
we can
simplify the problem by noting the periodicity of the function $\cos(ax)$
and  $\sin(a x)$. For example,
\begin{IEEEeqnarray}{rCl}
    &\int_{0}^{\infty}&\sin(a x)\mathrm{e}^{\mathrm{i}(\lambda+\alpha) x}\mathrm{d} x=\sum_{j=0}^\infty \int_{2j\pi/a}^{2(j+1)\pi/a}
    \sin(a x)\mathrm{e}^{\mathrm{i}(\lambda+\alpha) x}\mathrm{d} x\nonumber\\
    & = & 
    \left(\sum_{j=0}^\infty \e^{\ui\left( \alpha+\lambda \right) 2 j \pi /
    a}\right)
    \frac{1}{2}\left(\frac{1}{\alpha+\lambda+a}-\frac{1}{\alpha+\lambda-a}\right)(-\mathrm{e}^{\mathrm{i}(\alpha+\lambda)\frac{2\pi}{a}}+1).
\end{IEEEeqnarray}
Clearly, convergence problem arises when $j \to +\infty$ since $\alpha$ has a
negative imaginary part. However, this is a difficulty resulting from
linearization, not inherent difficulties in the flow physics and its modelling. As
mentioned earlier, the interaction between shock and instability waves occurs
when the instability waves grow to be of sufficient amplitude, at which place
the nonlinear/linear saturation or even decay begins to take place. Considering
that both the instability and shock waves start to decay further
downstream of the jet~\citep{J.COHEN_jfm, 2013_annual_rev},  the effective interaction only takes place within a
limited interval \adddd{spanning} several shock cells~\citep{1994Suda,Sources, X.D.Li}. In other words, the summation
only involves a finite number of terms and therefore the convergence problem
does not occur in realistic jets. In light of this, it is reasonable to only
focus on a limited number of shock cells in this paper, e.g. from the third to
the fifth \add{according to previous studies~\citep{1983Norm,Panda_standingwave,2014TAM,2017_jfm_sources}.} Furthermore, note that the right hand sides of~(\ref{equ:boundary condition 1}),~(\ref{equ:boundary condition 2}), and~(\ref{equ_sound_inhomogenous2}) all have the common factor  $\left(\sum\e^{\ui\left( \alpha+\lambda \right) 2 j
\pi / a}\right)$ after the Fourier \add{transform}. Hence they can be collected and accounted for later due to
linearity of the equation and the two boundary conditions. This means that we
only need to consider the interaction within one shock cell, as shown in
figure~\ref{fig:schematic}, and the total interaction field would be a simple
linear combination from a number of shock cells. In this way, the effective
integration interval when Fourier \add{transform} is applied to both the governing
equation and boundary conditions is limited to be within one shock cell.
\add{Physically, this corresponds to the effective sound generated by the interaction between the
instability waves and one single shock cell, similar to Powell's idea of treating the effective sound as that of monopoles.}

\addd{Note although we limit the integration interval to be within one shock cell, we do not imply that such an effective source is physically localized. The source term on the right side of (\ref{equ_sound_inhomogenous}) has a periodic nature by construction, but the bounds of integration may be across one or several shock cells due to the linearity of (\ref{equ_sound_inhomogenous}). This is equivalent to decomposing the problem into several sub-problems, each of which has an effective noise source within one shock cell. The overall sound is a linear combination of the solutions to these sub-problems. In the rest of this paper, we will focus on examining the characteristics of such an effective sound source \adddd{first and then discuss and compare the total sound from a number of these sources with experiments and numerical simulations.}}

\begin{figure}
		\centering	
		\includegraphics[width = 0.8\textwidth]{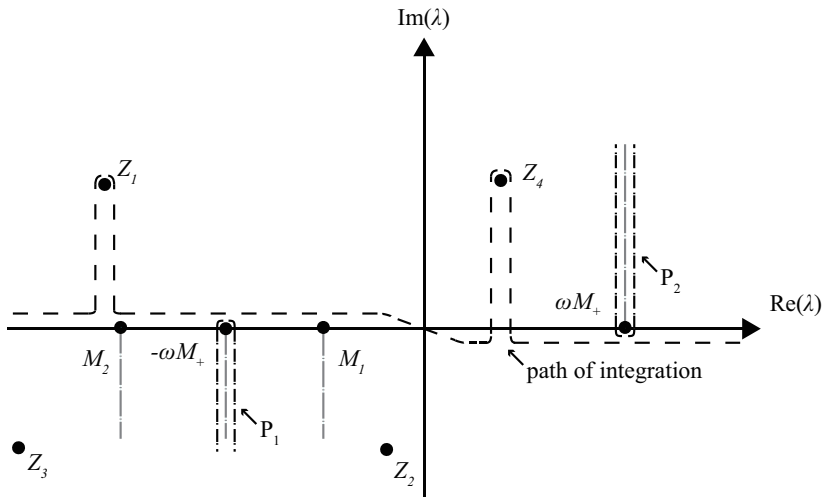}
		\caption{The branch points, branch cuts and integral path in complex $\lambda$ plane. $\mathrm{P}_1$ and $\mathrm{P}_2$ are the deformed integral paths when the saddle point approaches to the branch points $-\omega M_{+}$ and $\omega M_{+}$, \add{respectively}.}\label{fig:example3}
\end{figure}

Let us define
\begin{equation}
        \mathcal{I}_s(\lambda)=\int_{0}^{\frac{2\pi}{a}}\sin(a x)\mathrm{e}^{\mathrm{i}(\lambda+\alpha) x}\mathrm{d} x=\frac{1}{2}\big(\frac{1}{\alpha+\lambda+a}-\frac{1}{\alpha+\lambda-a}\big)\big(-\mathrm{e}^{\mathrm{i}(\alpha+\lambda)\frac{2\pi}{a}}+1\big),
        \label{integral_s_s}
\end{equation}
\begin{equation}
        \mathcal{I}_c(\lambda)=\int_{0}^{\frac{2\pi}{a}}\cos(a x)\mathrm{e}^{\mathrm{i}(\lambda+\alpha) x}\mathrm{d} x=\frac{\mathrm{i}}{2}\big(\frac{1}{\alpha+\lambda+a}+\frac{1}{\alpha+\lambda-a}\big)\big(-\mathrm{e}^{\mathrm{i}(\alpha+\lambda)\frac{2\pi}{a}}+1\big),
        \label{integral_s_c}
\end{equation}
with which the inhomogeneous equation can be written as
\begin{equation}
        \frac{\partial^{2}G_{-}}{\partial y^{2}}-\gamma_{-}^{2}G_{-}  = 
    2k_{1}\frac{A}{U}\bigg[\sinh(\zeta_{1}y)(B_{1}\mathcal{I}_{c}+B_{2}\mathcal{I}_{s})+ \sinh(\zeta_{2}y)(B_{3}\mathcal{I}_{c}+B_{4}\mathcal{I}_{s})\bigg],
\end{equation}
where
    \begin{equation}
        \gamma_{-}=\sqrt{\lambda^{2}-M_{-}^{2}(\omega+\lambda)^{2}}.
    \label{equ:mu-}
    \end{equation}
Equation~(\ref{equ:mu-}) is equivalent to
\begin{equation}
    \gamma_{-}=-\mathrm{i}\beta\sqrt{(\lambda-M_{1})(\lambda-M_{2})},
\end{equation}
where 
\begin{equation}
    M_{1}=\frac{-M_{-}\omega}{M_{-}+1},\quad M_{2}=\frac{-M_{-}\omega}{M_{-}-1}.
\end{equation}
The branch cuts \adddd{passing} $\lambda=M_{1}$ and $\lambda=M_{2}$ extend to the
lower half plane, as illustrated in figure~\ref{fig:example3}. The function
$G_{-}(\lambda,y)$ can be divided into two parts, a particular solution,
$G^{p}(\lambda,y)$,
and a complementary solution, $G^{c}(\lambda,y)$, i.e.
    \begin{equation}
        G_{-}(\lambda,y)=G^{p}(\lambda,y)+G^{c}(\lambda,y).
    \end{equation}
    The particular solution can be calculated analytically, i.e.
\begin{equation}
     G^{p}(\lambda,y)  = 
    2k_{1}\frac{A}{U}\bigg[\frac{\sinh(\zeta_{1}y)}{\zeta_{1}^{2}-\gamma_{-}^2}(B_{1}\mathcal{I}_{c}+B_{2}\mathcal{I}_{s})\nonumber\\
    + \frac{\sinh(\zeta_{2}y)}{\zeta_{2}^{2}-\gamma_{-}^2}(B_{3}\mathcal{I}_{c}+B_{4}\mathcal{I}_{s})\bigg ],
\end{equation}
and the complementary solution can be found to be
    \begin{equation}
        G^{c}(\lambda,y)=D_{2}(\lambda)\mathrm{e}^{\gamma_{-}y}+D_{3}(\lambda)\mathrm{e}^{-\gamma_{-}y},
    \end{equation}
    where $D_{2}$ and $D_{3}$ are two undetermined coefficients. 
    
    Applying \add{the}
    Fourier \add{transform} to the two boundary conditions, we can solve the
    undetermined coefficients. For the antisymmetric mode, we obtain
\begin{equation}
\begin{split}
D_{1}(\lambda) &=  \frac{A\mathrm{e}^{\frac{1}{2}\gamma_{+}}}{2U\eta M_{+}^{2}}\bigg[\gamma_{-}\coth(\frac{\gamma_{-}}{2})\mathcal{I}_{c}\sin(\pm\frac{1}{2}a\beta )\beta\big[\omega\big(2k_{1}m_{-}\cosh(\frac{1}{2}m_{-})\pm m_{+}\mathrm{e}^{\mp\frac{1}{2}m_{+}}\big)\\ 
& - \frac{1}{\omega}m_{+}\frac{M_{-}^{2}}{M_{+}^{2}}a^{2}\mathrm{e}^{-\frac{1}{2}m_{+}}\big]+M_{-}^{2}(\omega+\lambda)\bigg( \gamma_{-}\coth(\frac{\gamma_{-}}{2})\frac{2U}{A}G^{p}(\pm\frac{1}{2})\\
& - \frac{2U}{A}{G^{p}}^{\prime}(\pm\frac{1}{2})-i a \beta \mathcal{I}_{s}[b_{1}\sin(\pm\frac{1}{2}a\beta )+c\beta\cos(\frac{1}{2}a\beta )]\\
&-\mathcal{I}_{c}[b_{2}\sin(\pm\frac{1}{2}a\beta )-c\alpha\cos(\frac{1}{2}a\beta )]\bigg)\bigg],
\label{equ:resultD1_1_2}
\end{split}
\end{equation}
\begin{equation}
\begin{split}
D_{2}(\lambda) &=- \frac{A}{4\eta U}\bigg[\frac{\omega+\lambda}{\omega}\mathcal{I}_{c}\gamma_{+}\beta\sin(\pm\frac{1}{2}a\beta )\frac{1}{M_{+}^{2}}\big[\omega\big(2k_{1}m_{-}\cosh(\frac{1}{2}m_{-})\pm m_{+}\mathrm{e}^{\mp\frac{1}{2}m_{+}}\big)\\ 
&-\frac{1}{\omega}m_{+}\frac{M_{-}^{2}}{M_{+}^{2}}a^{2}\mathrm{e}^{-\frac{1}{2}m_{+}}\big]+\frac{M_{-}^{2}}{M_{+}^{2}}\frac{(\omega+\lambda)^{2}}{\omega}\gamma_{+}\frac{2U}{A}G^{p}(\pm\frac{1}{2})\\
&+\frac{2U}{A}{G^{p}}^{\prime}(\pm\frac{1}{2})+\mathrm{i}\omega a \beta \mathcal{I}_{s}[b_{1}\sin(\pm\frac{1}{2}a\beta )+c\beta\cos(\frac{1}{2}a\beta )]\\
&+\omega\mathcal{I}_{c}[b_{2}\sin(\pm\frac{1}{2}a\beta )-c\alpha\cos(\frac{1}{2}a\beta )]\bigg],
\end{split}
\label{equ:resultD2_1}
\end{equation}
and $D_{4}(\lambda)=-D_{1}(\lambda)$, $D_{3}(\lambda)=-D_{2}(\lambda)$, where $\eta(\lambda)$ is
\begin{equation}
    \eta(\lambda)=\omega{\rm c o t h}(\frac{1}{2}\gamma_{-})\gamma_{-}+\frac{1}{\omega}(\omega+\lambda)^{2}\gamma_{+}\frac{M_{-}^{2}}{M_{+}^{2}}.
\end{equation}
	It is straightforward to verify that $\lambda=-\alpha$ is a simple zero for $\eta(\lambda)$. In fact, $\eta(-\alpha)=0$ corresponds to the dispersion relation~(\ref{dispersion_anti}).
Besides, $\lambda=-\alpha$ is a simple                     
zero for $\mathcal{I}_s(\lambda)$
and a second-order zero for $\mathcal{I}_c(\lambda)$, so $\lambda=-\alpha$ is not a pole for $D_{1}(\lambda)$ and $D_{2}(\lambda)$. \adddd{It is found that $D_{1}(\lambda)$ and $D_{2}(\lambda)$ have four poles, which are}
\begin{gather}
    Z_{1,2}=\dfrac{\omega M_{-}^2\pm\sqrt{M_{-}^{2}\omega^{2}+(1-M_{-}^{2})\zeta_{1}^{2}}}{1-M_{-}^2},\\
    Z_{3,4}=\dfrac{\omega M_{-}^2\pm\sqrt{M_{-}^{2}\omega^{2}+(1-M_{-}^{2})\zeta_{2}^{2}}}{1-M_{-}^2}.
\end{gather}
The following inverse Fourier \add{transform}
\begin{equation}
        g_{+}(x,y)=\frac{1}{2\pi}\int_{-\infty}^{+\infty}D_{1}(\lambda)\mathrm{e}^{-(\mathrm{i}\lambda x+\gamma_{+}y)}\mathrm{d}\lambda
        \label{equ:numerical integration}
\end{equation}
yields $g_{+}$. 
The integration path is near the real axis
of $\lambda$, as illustrated in figure~\ref{fig:example3}. The integration path is \adddd{indented} to pass above the poles at $\lambda=Z_{1,4}$, as illustrated in figure~\ref{fig:example3}, in accordance with the causality argument~\citep{Briggs}. Because the real part of $\gamma_{+}$ should be positive when $|\lambda|\rightarrow \infty$ \adddd{along the integration path}, the branch
cuts of  $\gamma_{+}$  \adddd{passing the} branch points $\lambda=\pm \omega M_{+}$ are chosen to extend to the upper and lower half plane, respectively, as shown in figure~\ref{fig:example3}. 
The branch points of $\gamma_{-}$, i.e. $M_{1}$ and $M_{2}$, are on the 
negative real $\lambda$ axis. The branch cuts are chosen to extend down to the lower half plane so as not to  cross the integration path. Using the
steepest descent method, and noting that the saddle point is located at
$\lambda=-M_{+}\omega{\rm cos}\theta$, where $\theta={\rm a r c t a n}(y/x)$ representing the observer angle,
we can express $g_{+}$ as a function of radial distance $r$ and 
$\theta$ in the far field~($r\gg1$), i.e.
    \begin{equation}
        g_{+}(r,\theta)=\frac{\sqrt{M_{+}\omega}}{\sqrt{2\pi}}D_{1}(-M_{+}\omega{\rm c o s} \theta){\rm s i n} \theta \frac{\mathrm{e}^{\mathrm{i}\omega(M_{+}r-\pi/4)}}{\sqrt{r}}+O(r^{-3/2}),
        \label{steepest_decent_way}
    \end{equation}
\adddd{and with (\ref{20}) and (\ref{potential_outside}),  the corresponding pressure perturbation (nondimensionalized by $\sqrt{2/\pi}\rho_{0+}U^2$) can be expressed as}
   \begin{equation}
        p_{+}(r,\theta)=\mathrm{i}\sqrt{M_{+}}\omega^{\frac{3}{2}}D_{1}(-M_{+}\omega{\rm c o s} \theta){\rm s i n} \theta \frac{\mathrm{e}^{\mathrm{i}\omega(M_{+}r-t-\pi/4)}}{\sqrt{r}}+O(r^{-3/2}).
        \label{steepest_decent_way_p}
    \end{equation}

Note that the saddle
point moves between $-\omega M_{+}$ and $\omega M_{+}$ as $\theta$ changes from 0 to $\pi$, and the integral path is
forbidden to pass through the branch cut. So when $\theta=\pi$ and $0$, the
corresponding integral path is deformed along the branch cut and wraps the
branch point as shown by $\mathrm{P}_1$ and $\mathrm{P}_2$ in
figure~\ref{fig:example3}, respectively. It is similar when the steepest decent path passes through the branch cut at $M_{1}$ and $M_{2}$, in which case the 
integral path needs to be adjusted to avoid the branch cut. Note when the poles cross the steepest decent path, care must be taken regarding the residue contribution.

As can be seen from equation~(\ref{steepest_decent_way}), when the observer angle $\theta=\pi$, the potential function $g_{+}(r,\theta)$ reduces to a high order term $O(r^{-3/2})$ if $|D_{1}(-\omega M_{+}\cos\theta)|$ is bounded. This implies that sound waves propagating in this direction decays rapidly and nearly vanishes in the far field as $r\rightarrow \infty$. \adddd{This leads to an important feature of the sound directivity that will become clear in the rest of this paper.}

\section{Results and discussion}
\label{section:results and discussion}
\begin{figure}
		\centering
		\includegraphics[width = 0.55\textwidth]{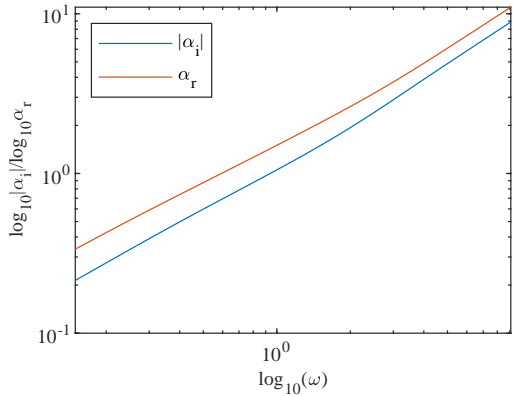}
		\caption{The real and imaginary parts of the solution of the spatial wavenumber $\alpha$ calculated from the dispersion relation~(\ref{dispersion_anti}). The antisymmetric mode is considered. Note that the imaginary part of the wavenumber $\alpha$ is negative, and its absolute value is plotted instead. }\label{fig:example4}
	\end{figure}

The sound field due to the interaction between shock and instability waves is shown in this section. In the linear stability analysis, the spatial wavenumber $\alpha$ is the central parameter determining the characteristics of the instability waves. 
The dispersion relation calculated from~(\ref{dispersion_anti}) is shown in figure~\ref{fig:example4}. The antisymmetric mode is considered. We see that both the wavenumber and growth rates increase as $\omega$ increases. These are well-established results in the linear stability analysis. Due to a negative 
imaginary part, the instability waves grow exponentially downstream the jet flow, and subsequently interact with shock cell structures.  

In what follows, we first examine the screech frequency prediction using \adddd{the present} model. Sound propagating at the observer angle $\theta=150^\circ$ is used to verify the far-field approximation before the directivity patterns of the fundamental tone and its harmonics are shown. Finally, we examine the near-field pressure fluctuation and discuss the noise generation  mechanism.
%
\subsection{The screech frequency}
\label{subsection:frequency_prediction}
\citet{1953Powell} proposed a model to predict the screech frequency by assuming a constructive interference in the upstream direction $\theta=180^\circ$. Following Powell's idea, the screech frequency and its harmonics can be calculated using \adddd{the present} model. Note similar frequency predictions have been investigated in earlier studies, nevertheless, it is included here as a validation of the model. The shock cell spacing $\add{s}$  satisfies
\begin{equation}
    \add{s}/D=\frac{2\pi}{a}=2\sqrt{M_{-}^{2}-1}.
    \label{equ:shock space}
\end{equation}
\citet{Tam_shock_space} showed that in the jet flow from a rectangular nozzle, the shock cell spacing satisfied
\begin{equation}
    \add{s}/D=\dfrac{2\sqrt{M_{-}^{2}-1}}{\sqrt{1+\bigg(\dfrac{D}{b}\bigg)^{2}}},
    \label{tam's prediction}
\end{equation}
where $b$ is the width of the \adddd{fully-expanded} jet flow. When the rectangular jet is of high aspect ratio~$(D/b\ll1)$, we can see that~(\ref{tam's prediction}) reduces to~(\ref{equ:shock space}). \add{Note that (\ref{equ:shock space})
may not be able to predict the shock spacing accurately when the magnitude of overexpansion/underexpansion increases~\citep{X.D.Li_shock}, while a correct representation of shock structures plays a dominant role in predicting the screech frequency, especially in circular jets~\citep{A1A2modes}. Nevertheless, in this planar model, the jet is assumed to be slightly imperfectly-expanded, and previous studies~\citep{Tam_shock_space} showed good agreement between the prediction and the experimental data. Therefore, we choose~(\ref{equ:shock space}) to predict the shock spacing.}

\begin{figure}
	\centering
	\label{1frequency spectrum}
	\includegraphics[width = 0.55\textwidth]{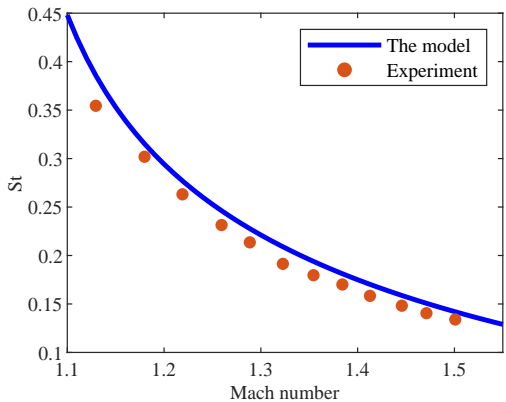}
	\caption{\addd{Comparison of the fundamental screech frequency between  Powell's experiment~\citep{1953Powell} and the model's prediction.}}
	\label{frequency_check}
	\end{figure}

The convection velocity of instability waves is widely believed to be proportional to the velocity of the fully-expanded jet flow, i.e.
\begin{equation}
    U_{c}=\kappa U_{-},
    \label{equ:convective velocity}
\end{equation}
\add{where $\kappa$ is usually taken to be $0.7$.}
Equations~(\ref{equ_screech f}),~(\ref{equ:shock space}), and~(\ref{equ:convective velocity}) can be combined to yield the nondimensionalized angular frequency of the sound wave
\begin{equation}
    \label{screech_omega}
    \omega=\frac{a m \kappa }{M_c +1},
\end{equation}
where $m=1, 2, 3, 4...$ \adddd{correspond to} the fundamental frequency, the first, second and higher harmonics, respectively. \addd{Considering that $M_c=U_c/a_\infty$, where $a_\infty$ is the speed of sound in the free stream. \adddd{For a cold jet, $M_c$ can be calculated by} }
\addd{\begin{equation}
    M_c=\frac{\kappa M_{-}}{\sqrt{1+\frac{\gamma-1}{2}M_{-}^2}}.
    \label{3.5}
\end{equation}
\adddd{With~(\ref{3.5}), $\omega$ can be readily calculated via~(\ref{screech_omega}). This formula is consistent with that derived by~\citet{Tam_shock_space}.}} A comparison between the measured fundamental screech frequency by~\citet{1953Powell}
and that predicted by~(\ref{screech_omega}) is shown in figure~\ref{frequency_check}. \adddd{As can be seen, good agreement is achieved. Equations (\ref{screech_omega}) and (\ref{3.5}) show that when the jet operating condition is known, the screech frequency can be readily calculated.}
The operating conditions and frequencies calculated in this way, as shown in table~\ref{tab:kd}, will be used in following sections.

\add{
Note that this paper does not attempt to model the entire feedback loop, and the reason we include a frequency prediction is mainly to validate the model. \adddd{Thus, although the feedback theory proposed by Powell is used, it is only used to predict the frequency using (\ref{screech_omega}). Our focus in this paper is to model the interaction between the shock and instablity waves.} It is worth noting that a new  feedback mechanism for circular jets has been proposed in a number of recent papers~\citep{Gojon_modeStaging,edgington_modestaging,Xiang-Ru_2020,NewC_1}. The present paper, however, focuses on a 2D jet, and it is not yet clear what role the guided jet modes play in this case. Besides, even for circular jets, it is known that the convection velocity of the upstream-travelling jet guided mode is very close to the speed of sound. Therefore, it can be expected little change in the frequency prediction would occur even if the guided jet mode is taken as the closure mechanism.}

\begin{table}
  \begin{center}
\def~{\hphantom{0}}
  \begin{tabular}{lccc}
      $M_{-}$  & $\omega_{1}$   &   $\omega_{2}$ & $\omega_{3}$ \\[3pt]
       1.5   &\addd{1.05} & \addd{2.10} & \addd{3.15}\\
       1.3   & \addd{1.48} &\addd{2.86} &\addd{4.45}\\
       1.2  & \addd{1.91} & \addd{3.81} &\addd{5.73}\\
  \end{tabular}
  \caption{The operating conditions and calculated frequencies of the screech tones, where $\omega_{1}$ denotes the angular frequency of the fundamental tone, $\omega_{2}$ its first harmonic, and $\omega_{3}$ the second harmonic.}
  \label{tab:kd}
  \end{center}
\end{table}

\subsection{Directivity of the sound field }
\label{subsection_directivity}

\add{The distinct directivity pattern of jet screech is perhaps one of its most important features and has been well-reported in various experiments. In this section, we aim to predict the noise directivity using the model developed in section 2. As mentioned in section 1, this paper concerns the acoustic emission due to the interaction between shock and instability waves, and therefore does not consider the entire feedback process of screech. However, the existence of the feedback would not \adddd{alter} the fact that the noise is generated due to the shock-instability interaction. Therefore, \adddd{provided the screech frequency is specified, the present model can be used to compare with the screech directivity.}}

\begin{figure}
	\centering
	\includegraphics[width = 0.55\textwidth]{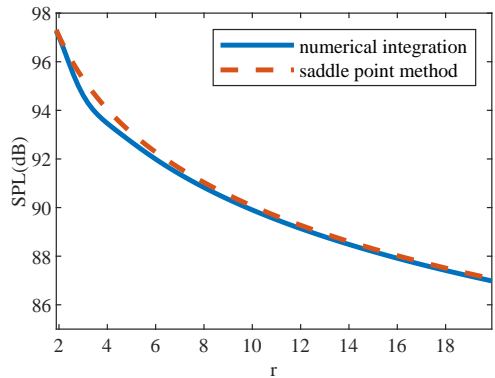}
	\caption{\addd{Comparison of the SPL from the far-field approximation and that from the numerical integration~($\theta=150^\circ$). The Mach number of the fully-expanded jet is 1.5. The origin $r=0$ represents the beginning of the first shock cell, and $r$ is the nondimensionalized radial distance.}}
		\label{sound_outside}
	\end{figure}

Before used to study the directivity of the resulting sound, equation~(\ref{steepest_decent_way_p}) is verified by numerically integrating~(\ref{equ:numerical integration}) at $\theta=150^\circ$. The comparison between the prediction using~(\ref{steepest_decent_way_p}) and~(\ref{equ:numerical integration}) is shown in figure~\ref{sound_outside}. The SPL is defined to be
\begin{equation}
    \mathrm{SPL}=10\log_{10}\frac{|p_{+}|^2}{|p_{\mathrm{ref}}|^2},
    \label{equ:SPL}
\end{equation}
\adddd{where the reference pressure $p_\mathrm{ref}=2\times 10^{-5}$.} We see that good agreement is achieved when $r$ is beyond $5$, where the difference between two methods is within 1 dB. When $r\geq 20 $, the difference reduces to 0.2 dB. \adddd{Consequently $r=5$, in which case $kr\approx7.9$, may be used to approximately separate the acoustic near and far field~\citep{krgg1}.}

 Whether the sound source of jet screech is spatially localized or distributed along the jet flow is still open to debate. As mentioned above, Powell proposed the monopole array theory to predict the directivity patterns. Many other researchers~\citep{S.kaji,1997JFM_Raman,Sources} also found that the fundamental screech tone emitted from several shock cells downstream the jet flow. However, other researchers~\citep{1997_POF_SHWalker,2017_jfm_sources} observed that the screech was produced from a particular shock cell, e.g. the third or fourth one. In addition, “shock-clapping”~\citep{1994Suda} and “shock leakage”~\citep{2003Suzeki} were observed at the third and fourth shock cells downstream the jet, which were suggested to be the source of screech, particularly for higher harmonics~\citep{2020_sound_sources}. \add{This suggests that the number of shock cells that need to be included is not clear. \adddd{However}, it would be interesting to compare and contrast the noise directivity patterns due to the interaction between the instability waves and various numbers of shock cells.} Therefore, in what follows we will examine the directivity patterns due to the interaction between the instability wave and a single shock cell, \adddd{and then discuss that from several shock cells.}
 


It is known that in realistic jets the instability waves exhibit a characteristic structure of wave packets~\citep{2005Lele,2013_annual_rev,2019Wong}. The amplitude of the instability waves varies slowly within one wavelength~\citep{2006_jfm_wavepackets}, while the whole wave packet shows a Gaussian envelope~\citep{2011_jsv_wavepackets}, \add{or more precisely an exponentially-modified Gaussian envelope as demonstrated by a recent work~\citep{exponetially_Gaussian}}. We see that the local growth rate within a wavelength is varying, but the effects of the local growth rate on the sound characteristics are not clear \adddd{across the wave packet}. In this paper, we will also examine the effects of the local growth rate by showing results \add{with various values of $\alpha_{\mathrm{i}}$}.

\subsubsection{Directivity pattern of \add{sound due to one-cell interaction}}

When the \adddd{fully-expanded Mach number} is given, the directivity patterns of the fundamental tone and its harmonics can be calculated from~(\ref{steepest_decent_way_p}). The operating condition is shown in table~\ref{tab:kd}. Note that  the wavenumber is obtained from~(\ref{dispersion_anti}), as we  consider the antisymmetric mode.
 	\begin{figure}
	\centering
	\includegraphics[width = 1.0\textwidth]{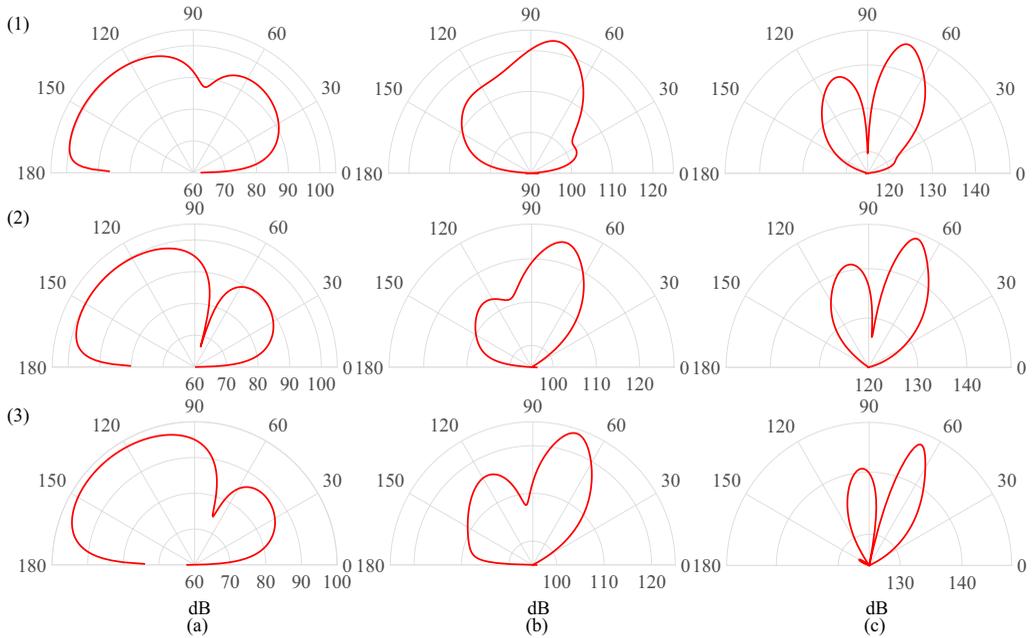}
	\caption{\addd{The directivity of sound in the far field obtained by~(\ref{steepest_decent_way_p}). $r $ is fixed to be 1. Labels (1), (2), (3) represent the results for a fully-expanded jet Mach number of 1.5, 1.3, and 1.2, respectively. Columns (a), (b), (c) are the results of the fundamental tone, the first and second harmonics, respectively. The antisymmetric mode of instability waves is taken, and the imaginary part of wavenumber $\alpha_{\mathrm{i}}\neq 0$. }\adddd{In addition, $\mathcal{U}$ in~(\ref{equ:A}) is taken to be 1.}}
		\label{n=1_consider}
	\end{figure}
The directivity patterns of the fundamental tone and its first two harmonics under three different operating conditions are shown in figure~\ref{n=1_consider}. The SPL is defined by~(\ref{equ:SPL}). Labels (1), (2), (3) represent the results for the fully-expanded jet Mach number of 1.5, 1.3, and 1.2, respectively. Columns (a), (b), (c) are the results of the fundamental tone, the first and second harmonics, respectively. \addd{From figure~\ref{n=1_consider}, it is clear that the \addd{effective} directivity of the fundamental tone due to a single shock cell is not that of a monopole. Instead, it consists of two
lobes. One primary lobe radiates upstream, while the other radiates downstream with a weaker intensity. Although this represents the effective directivity due to one shock-cell interaction, we see that it possesses some inherent directivity that resembles the total sound field measured in experiments. For example, the effective directivity of the fundamental tones accords with} both numerical~\citep{numerical_directivity_of_rectangular} and experimental~\citep{1997_POF_SHWalker,Sources} results for rectangular jets of high aspect ratios, \add{where both an upstream lobe and a downstream lobe of similar intensities appear. However, a precise match of every lobe position may not be achieved; this is expected because the prediction is only from one shock-cell interaction, whereas the numerical and experimental results are from the entire screeching jet.} Moreover, in all three cases, the fundamental tone is reinforced at the upstream direction, but drops quickly as $\theta$ approaches $180^\circ$. In Kerschen's original paper~\citep{1995Kerchen}, the directivity pattern reaches its maximum at $48^\circ$ with only one lobe in the downstream direction. Little sound radiates in the upstream direction. Our present model shows that when a two-dimensional vortex sheet and a more realistic shock structure are considered, the screech
directivity from a single shock cell has a major radiation lobe in the upstream direction. This result shows a better \adddd{qualitative} agreement with experiments. It is also interesting to note that the maximal radiation angles shown in figure 7 appear to depend on the fully-expanded jet Mach number. A similar tendency was also reported in earlier experiments for round jets~\citep{1992Powell}. 

Another important result is that there is a large lobe perpendicular to the jet flow for the first harmonic in all three conditions, \adddd{which again resembles the total noise directivity measured in experiments}~\citep{1953Powell, 1997_POF_SHWalker, 2020_sound_sources}. Note that the peak radiation angle was reported to be not exactly at $90^\circ$, but slightly towards the downstream direction~\citep{numerical_directivity_of_rectangular, M.Kandula, 2020_sound_sources}, \adddd{which is similar to the prediction by the present model}. For the second harmonic,~\add{it was experimentally observed that the \add{effective} directivity pattern showed two lobes, one directed slightly upstream, and one downstream~\citep{Sources}; this feature is \adddd{consistent} with the prediction.  In addition, we find that in a recent experiment conducted by~\citet{2020_sound_sources}, the main radiation angle for the second harmonic is between $40^\circ$ and $110^\circ$, while little radiation appeared around $90^\circ$. We see from figure~\ref{n=1_consider} that this \adddd{is also reflected} in the prediction.} 

\addd{From figure~\ref{n=1_consider}, It is straightforward to see that the effective noise directivity due to the interaction between the instability waves and one shock cell is not of the monopole type, but shows an intrinsic shape that is close to that of the overall screech directivity. This shows that the unique directivity of jet screech is not caused by pure interference between an array of monopoles as assumed by Powell. Therefore, to properly model and understand the directivity of screech, one has to use quantitative models such as the one developed in this paper. However, it is worth noting that this model does not imply that the screech source is localized as a “single” source, what we show is just the effective noise directivity due to interactions between the instability waves and a single shock cell.}

\begin{figure}
	\centering
	\includegraphics[width = 0.75\textwidth]{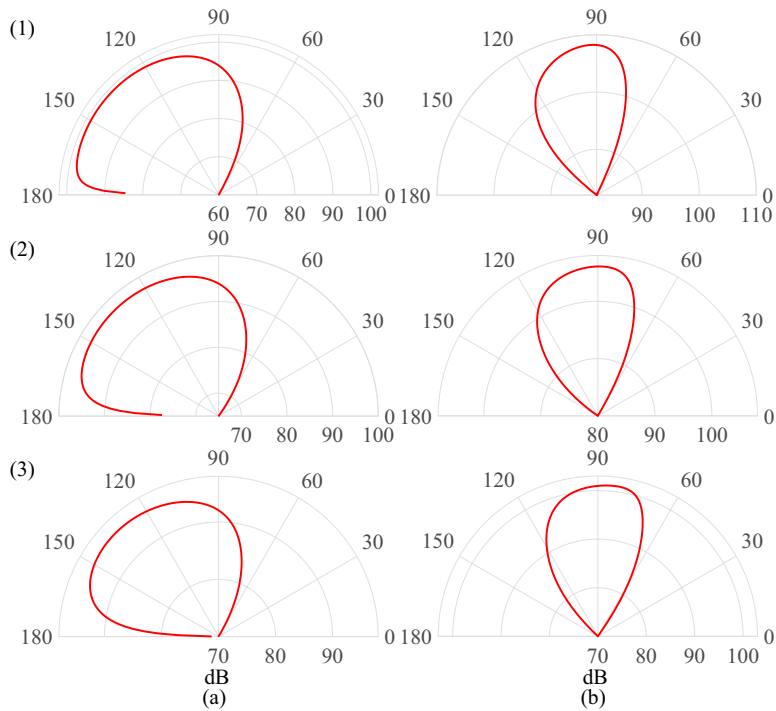}
	\caption{\addd{The directivity of sound in the far field obtained by~(\ref{steepest_decent_way_p}). $r $ is fixed to be 1. Labels (1), (2), (3) represent the results for a fully-expanded jet Mach number of 1.5, 1.3 and 1.2, respectively. Columns (a), (b) are the results of the fundamental tone and the first harmonic, respectively. The antisymmetric mode of instability waves is taken, and the imaginary part of wavenumber $\alpha_{\mathrm{i}}= 0$. }\adddd{In addition, $\mathcal{U}$ in~(\ref{equ:A}) is taken to be 1.}}
		\label{n=1_noconsider}
	\end{figure}

\add{To examine the effects of the local growth rate of the instability waves on sound generation, the result obtained using~(\ref{steepest_decent_way_p}) with only the real part of $\alpha$ considered is shown in figure~\ref{n=1_noconsider}.} The SPL is similarly defined by~(\ref{equ:SPL}). Only the fundamental tone and its first harmonic are presented to compare with those reported by Powell~\citep{19533Powell}. \addd{Considering that the original directivity results reported by~\citet{19533Powell} were presented in the form of schlieren photographs and were therefore not suitable for a direct comparison, only a qualitative comparison is presented.} 
As shown in figure~\ref{n=1_noconsider}, the fundamental tones in all three cases are only reinforced in the upstream direction, while in Powell's experiment, sound waves at the fundamental frequency can only be observed propagating upstream, as shown in figure 4 of the original paper~\citep{19533Powell}. \addd{The model prediction is in agreement with the experimental data}. In addition, a quick decay also occurs when the observer angle approaches $180^\circ$, which is similar to the case when the imaginary part of $\alpha$ is not zero. While for the first harmonic, a large  lobe perpendicular to the jet flow is predicted by our model. In Powell's experiment, when a reflector was placed, a downstream propagating sound wave~(as shown in figure~5 of the original paper~\citep{19533Powell}) twice of the fundamental frequency emerged. This implied that there was a strong beaming to the side of the jet flow, which is in good \adddd{qualitative} agreement with the prediction. 

\begin{figure}
	\centering
	\includegraphics[width = 0.75\textwidth]{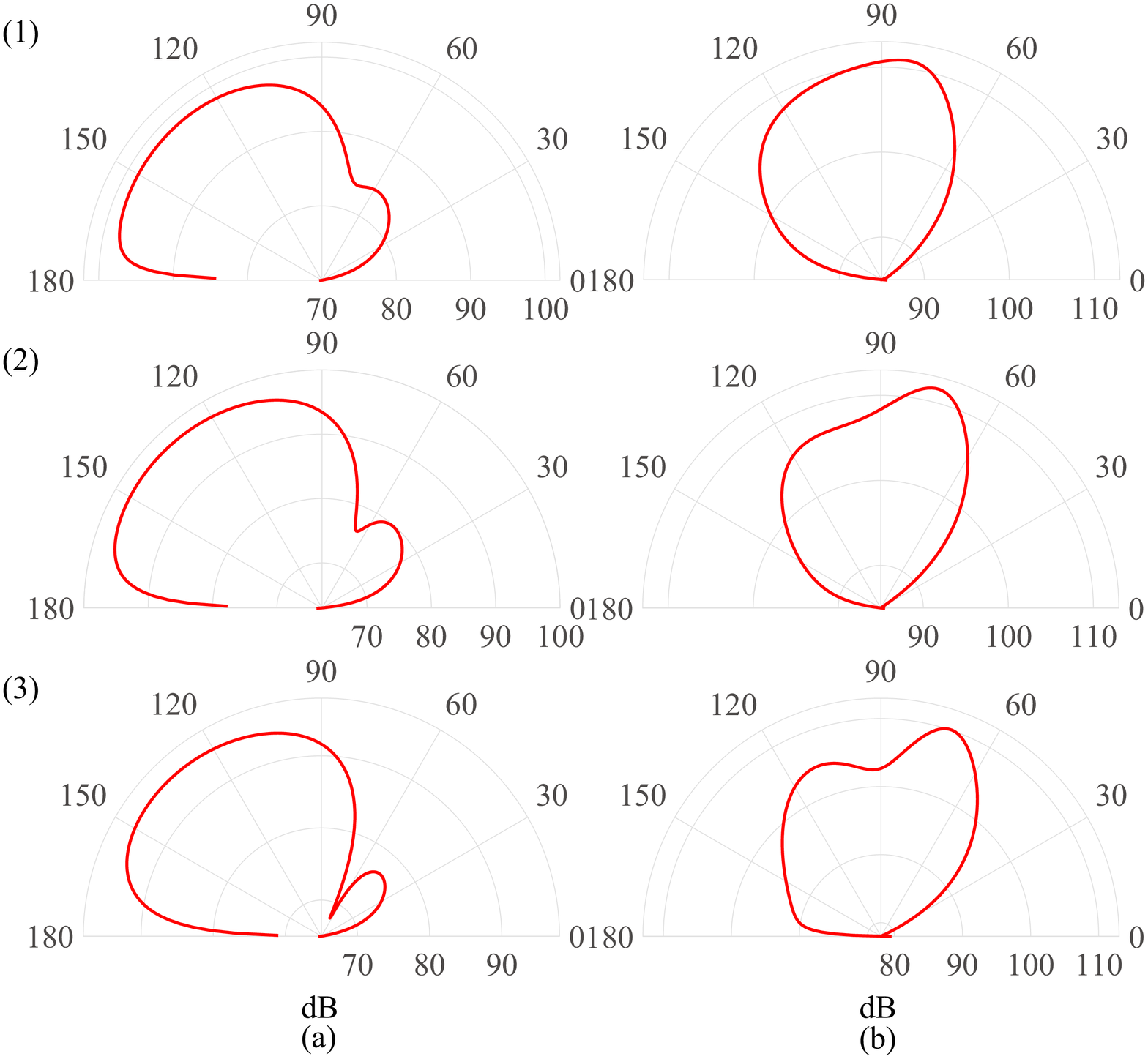}
	\caption{\addd{The directivity of sound in the far field obtained by~(\ref{steepest_decent_way_p}). $r $ is fixed to be 1. Labels (1), (2), (3) represent the results for a fully-expanded jet Mach number of 1.5, 1.3 and 1.2, respectively. Columns (a), (b) are the results of the fundamental tone and the first harmonic, respectively. The antisymmetric mode of instability waves is taken, and the imaginary part of wavenumber $\alpha_{\mathrm{i}}$ changes to $2/3$ of the original value.} \adddd{In addition, $\mathcal{U}$ in~(\ref{equ:A}) is taken to be 1.}}
		\label{n=2/3}
	\end{figure}

\add{To further investigate the effects of the local growth rate of the instability waves on directivity patterns, we change the imaginary part of $\alpha$  to $2/3$ of its original value. The resulting directivity patterns are shown in figure~\ref{n=2/3}. As can be seen, the fundamental tones in all three cases radiate primarily to the upstream direction, while small lobes appear downstream of the jet flow. Compared with the results shown in figure~\ref{n=1_consider} these lobes are both thinner and weaker, whereas in the case of $\alpha_{\rm{i}}=0$ there are no observable lobes downstream of the jet flow, as illustrated in figure~\ref{n=1_noconsider}(a). For the first harmonic, a large lobe appears perpendicular to the jet flow. Compared with the results shown in figure~\ref{n=1_consider}, the directivity patterns seem to shrink and move closer to $90^\circ$ to the jet. In particular, two small lobes appearing in figure~\ref{n=1_consider}(3b) appear to collapse to a single wide lobe, as shown in figure~\ref{n=2/3}(3b).}

The directivity pattern when a \add{positive} imaginary part of $\alpha$ is used can be similarly studied. It can be shown that little change  occurs when $\alpha$ is replaced by $\alpha^{*}$~($\alpha^{*}_{i}=-\alpha_{i}$). So we will omit a \adddd{repetitive} discussion for brevity. Figures~\ref{n=1_consider},~\ref{n=1_noconsider}, and~\ref{n=2/3} show that the directivity pattern depends on the local growth rate of the instability wave, and a change in the imaginary part of $\alpha$ would lead to a corresponding  change in the resulting directivity pattern.
As mentioned at the beginning of  section~\ref{subsection_directivity}, it is not clear exactly where the interaction between instability and shock waves occurs. The amplitude of instability waves may have experienced a growth, a saturation, or even decay before reaching the point of interaction. Comparing figures~\ref{n=1_consider},~\ref{n=1_noconsider}, and~\ref{n=2/3}, we see that the noise directivity is very sensitive to the local growth rate, and this may be used to explain the discrepancies observed across different experiments. For example, Powell's original results showed a clearly dominant radiation only in the upstream direction and a strong $90^\circ$ radiation at the first harmonic. This could be explained well if the local instability waves experience a saturation. On the other hand,~\citet{1997_POF_SHWalker},~\citet{numerical_directivity_of_rectangular_2}, and~\citet{GJWU}  showed that two lobes could be observed at the fundamental frequency, and a relatively weak radiation at $90^\circ$ for the \addd{first} harmonic. This may be explained if the instability waves are in a growth or decay stage at the effective point of interaction.

\subsubsection{Directivity patterns from several shock cells}
As mentioned at the beginning of section~\ref{subsection_directivity}, some researchers report that screech appears to originate from several shock cells downstream the jet flow, for example, from the second  to the fourth shock cell~\citep{1994Suda,Sources}. \addd{Since our model \adddd{can include multiple shock structures, we can study and compare the sound field produced by the interaction between instability waves and several shock cells with simulations and experiments}}. \addd{Note that to precisely predict the screech amplitude, it is likely that every stage of the feedback loop needs to be considered~\citep{absolute_instability} and the nonlinearity that is inevitable within the loop needs to be included. However, in this paper, we only study the shock-instability interaction and use a linear model ($\mathcal{U}$=1 in~(\ref{equ:A})). Therefore, the prediction would not be able to match the data in terms of the absolute amplitude. However, we can still plot the predictions and the numerical or experimental data in one figure and focus instead on comparing the shapes of the directivity pattern.  The SPL of the model prediction is again defined by~(\ref{equ:SPL}), but rescaled according to the experimental or numerical data.} \addd{The predictions of the monopole array theory are also included \adddd{for} comparison.} 

\addd{The results are first compared with the study by~\citet{GJWU}, where LES simulations were conducted and well validated against the experimental data reported by~\citet{Wu_experiment_1} and~\citet{Wu_experiment_2}. In both the experiment and the numerical simulation, a rectangular nozzle with an aspect ratio of 4:1 was used. The designed Mach number of the nozzle was 1.44, while the Mach number of the fully-expanded jet flow was 1.69.} It was stated by~\citet{GJWU} that the measured directivity patterns resulted from the interference among spatially distributed sources. Therefore, multiple shock cells are included in our model to facilitate a comparison. Considering that the amplitude of instability waves shows a Gaussian~\citep{2011_jsv_wavepackets}, \add{or more precisely exponentially-modified Gaussian~\citep{exponetially_Gaussian}} intensity distribution downstream of the jet flow, in this paper \add{we use three shock cells, the interaction between which and the instability waves leads to different effective source strengths.} The middle cell is chosen to have the maximal strength. In front of the middle cell, the instability waves still grow and have not reached the maximum intensity, while after that the instability waves begin to decay but are still of sufficient intensity to generate sound. \addd{Thus, the relative strengths of the three interactions are assumed to be 0.45, 1, 0.7, respectively.} Similar assumptions have been also made by~\citet{1983Norm} and \citet{ numerical_directivity_of_rectangular}. \addd{It was known that the effect of varying source strengths on the directivity of the fundamental and the first harmonic was unimportant with regard to the principal lobe, and was appreciable only in the secondary or minor lobes~\citep{M.Kandula}. }In light of linearity, in what follows we first calculate the sound by one shock cell using our model, and then combine the other two with a spatial phase difference $ \mathrm{e}^{\mathrm{i}\left(\lambda_{0}+\mathrm{real} (\alpha)\right)2 \pi/a}$, where $\lambda_{0}=-\omega M_{+}\cos\theta$. 
 	\begin{figure}
	\centering
	\includegraphics[width = 0.75\textwidth]{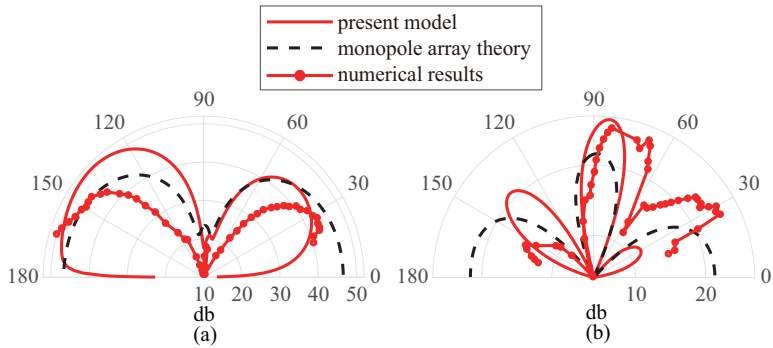}
	\caption{\addd{The comparison between the numerical results~\citep{GJWU}, the present model, and the monopole array theory~\citep{19533Powell}. The Mach number of the fully-expanded jet flow is 1.69. The red solid line denotes the model prediction, the red line with markers the numerical data, and the black dashed line represents the prediction of the monopole array theory. Columns (a) and (b) represent the fundamental tone and its first harmonic, respectively.}}
		\label{Norum_1.49}
	\end{figure}

    \begin{figure}
	\centering
	\includegraphics[width = 0.75\textwidth]{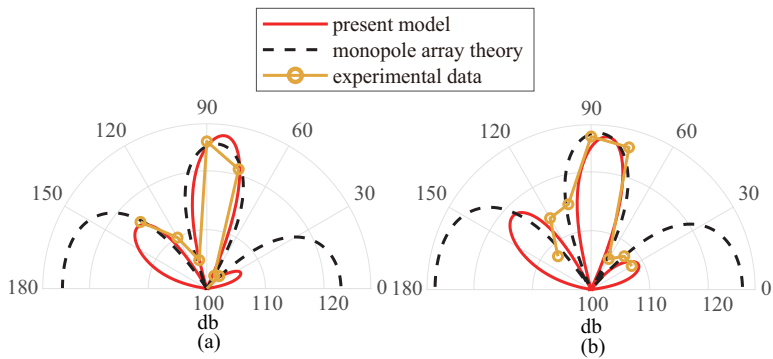}
	\caption{\addd{The comparison between the experimental data~\citep{farfield_rectangular}, the present model, and the monopole array theory~\citep{19533Powell}. The fully-expanded  Mach number for (a) and (b) is 1.5 and 1.6, respectively, while the designed Mach number of the nozzle is 1.35. The red solid line denotes the model prediction, the orange line with markers the experimental results, and the black dashed line represents the prediction of the monopole array theory. }}
		\label{Norum_1.19}
	\end{figure}

The result is shown in figure~\ref{Norum_1.49}, where 
columns (a) and (b) represent directivity patterns of the fundamental tone and the first harmonic, respectively. \addd{As can be seen, for the directivity pattern of the fundamental tone, the numerical data shows two lobes. The major lobe points to the upstream direction, while the other peaks \adddd{at} around $30^\circ$ with a slightly weaker intensity. Note that the radiation intensity seems to decrease as the observer angle approaches $0^\circ$. The results from the monopole array theory also show two lobes, which are of equal intensity and peak at $180^\circ$ and $0^\circ$, respectively. The weaker intensity of the downstream lobe appears not captured in the model. In addition, as $\theta$ approaches $0^\circ$, the monopole array theory predicts an increasingly large noise radiation, which contradicts the numerical data. On the other hand, we see that the present model predicts a similar major lobe in the upstream direction, and a weaker lobe in the downstream direction. The relative intensity and positions of the two lobes agree better with the numerical data than the monopole array theory. Moreover, the predicted acoustic radiation decreases quickly as the observer angle approaches $0^\circ$, which is in good agreement with the numerical data. Note however the maximum radiation angle of the downstream lobe appears at around $40^\circ$, slightly different from the numerical data. But considering the many assumptions made in the model, such deviation may be \adddd{deemed} acceptable.}

\addd{For the first harmonic, the numerical results exhibit three lobes. Two dominant lobes peak at $\theta=30^\circ$ and $\theta=83^\circ$, respectively, and one secondary lobe points to $155^\circ$. It is evident that a quick decrease occurs as the observer angle approaches $180^\circ$ and $0^\circ$, respectively. The results obtained by the monopole array theory also show three lobes, which are of the same intensity and peak at $180^\circ$, $88^\circ$, and $0^\circ$, respectively. The corresponding 
errors compared with the numerical data are $+25^\circ$, $+5^\circ$, and $-30^\circ$, respectively. Moreover, the monopole array theory predicts monotonously increasing acoustic radiations as the observer angle approaches $180^\circ$ or $0^\circ$, which is not able to match the numerical data. For the present model prediction, a narrow lobe peaks \adddd{at} around $84^{\circ}$ with a dominant intensity, and two weaker lobes appear \adddd{at} \addd{$\theta=31^\circ$} and \addd{$\theta=135^\circ$}, respectively. The corresponding differences compared with the numerical data are $+1^\circ$, $+1^\circ$, and $-20^\circ$, respectively, which is in more satisfactory agreement with the numerical results than those predicted by the monopole array theory. In addition, the rapid decay as the observer angle approaches $180^\circ$ and $0^\circ$ can be predicted well by this model. \adddd{Note however the present model cannot correctly predict the relative amplitude of the upstream and downstream lobes. The reason is not yet clear.} In summary, although \adddd{both} models are not capable of predicting the screech amplitude, the present model  shows a more satisfactory agreement with the numerical data in terms of the peak angle. In addition, unlike \adddd{the monopole array theory}, it can capture the rapid decay of the noise intensity as observer angles approach $0^\circ$ and $180^\circ$.} 

%

\addd{The results obtained by the present model are subsequently compared with~\citet{farfield_rectangular}, where a series of experiments were conducted \adddd{at} the NASA Langley Research Center. Several microphones were positioned on a circular arc from $\theta=30^\circ$ to $135^\circ$ at an increment of $15^\circ$. The aspect ratio and the designed Mach number of the rectangular nozzle \adddd{were} 3.7 and 1.35, respectively. Two sets of experimental data are chosen for comparison, of which the corresponding fully-expanded Mach numbers are 1.5 and 1.6, respectively. \adddd{Note that the available data only spans the observer angle between $30^\circ$ and $135^\circ$ at an increment of $15^\circ$. It is known that 
very weak noise is radiated in this range, and the peak radiation angles are likely to fall outside this range at the fundamental frequency. Therefore, for a robust comparison we only compare the first harmonic results, where it is known to radiate primarily at side angles.} The results predicted by the monopole array theory are also included for comparison. The number and relative strengths of the effective sources are kept the same as those used in figure 10. }

\addd{As can be seen in figure~\ref{Norum_1.19}, in the case of $M_{-}=1.5$, the experimental data shows a major lobe to the side of the jet, and another lobe  in the upstream direction with a slightly weaker intensity. In addition, note that in the downstream direction, the acoustic radiation becomes much weaker, as can be seen \adddd{at} the observer angle $\theta=30^\circ$. The results from the monopole array theory exhibit three lobes of nearly the same intensity, one in the upstream direction, one in the downstream direction, and another to the side of the jet. The agreement with the experimental data is good \adddd{when} $60^\circ<\theta<135^\circ$, but much less so in the downstream direction. The present model prediction also exhibits three lobes, one dominant lobe to the side of the jet, one weaker lobe \adddd{in} the upstream direction, and another lobe peaks at $25^\circ$ with a much weaker intensity. Both the position and the relative intensity of the three lobes agree well with the experimental data. }

\addd{In the case of $M_{-}=1.6$,  as can be seen in figure~\ref{Norum_1.19}(b), the experimental data shows a dominant lobe to the side of the jet, while another lobe appears slightly downstream with a weaker intensity. Note that a much weaker acoustic wave radiates to the downstream direction. The prediction obtained by the monopole array theory shows three lobes of the same intensity. \adddd{Again, the agreement with the experimental data is good} when $60^\circ < \theta < 120^\circ$, but much less satisfactory in the downstream direction. The present model prediction also exhibits three lobes, one dominant lobe to the side of the jet, one weaker lobe in the upstream direction, and another peaks at $30^\circ$ with a much weaker intensity. Both the position and the relative intensity of the latter two lobes agree well with the experimental data. In both the monopole array theory and the present model, it is difficult to drama a conclusion about the agreement when $\theta$ approaches $135^\circ$ due to the very sparse data points. Because the experimental results only cover an observer angle of $30^\circ$ to $135^\circ$, the directivity patterns as the observer angle approaches $180^\circ$ and $0^\circ$ \adddd{cannot} be examined. \adddd{However, the numerical results in figure 10 show a quick decay as the observer angle approaches $180^\circ$ and $0^\circ$. Similar phenomenon has been widely reported in numerous experiments for circular nozzles, such as those by~\citet{1983Norm} and \citet{1992Powell}. Such an important feature can be well captured by the present model, compared to earlier models.}} 

\addd{Note that the present model makes use of a number of linear approximations, for example, both the shock and instability waves are of small magnitude. \adddd{It is mentioned that supersonic jets may be regarded as weakly imperfectly-expanded when $|M_{-}^2-M_{1}^2|\le1$~\citep{tam_machwave,1985Tam}. It can be seen that both the LES simulation and the experiments satisfy this condition. Therefore, the present model may be used to compare with the two cases.} However, deviation may occur if intense shocks are involved. In such cases, we would not expect accurate predictions, but it may be possible that some important features of the \adddd{nonlinear screech} may still be captured by the linear model.}

\acc{In summary, in this section we show that the effective noise directivity due to the interaction between the instability waves and one shock cell has an intrinsic shape that is close to that of the overall screech directivity. In particular, figure 7(a) shows that the directivity for the fundamental tone has a major lobe in the upstream direction and  a minor lobe in the downstream direction. Comparing with figure 11(a), we see that  incorporating multiple shock-cell interactions results in two lobes that are both thinner and closer to $180^\circ$ and $0^\circ$, respectively. This can be expected, because multiple acoustic sources satisfying (3.4) would lead to constructive intereference near $180^\circ$ (and may or may not be so near $0^\circ$ depending on operating conditions). Therefore, the two radiation lobes after incorporating multiple effective sources would become thinner and closer to the jet centerline. Similar trends can be observed for the first harmonic by comparing figures 7 and 11 }

\subsection{Near-field pressure and noise generation mechanism}
\begin{figure}
    \centering
    \includegraphics[width = 0.6\textwidth]{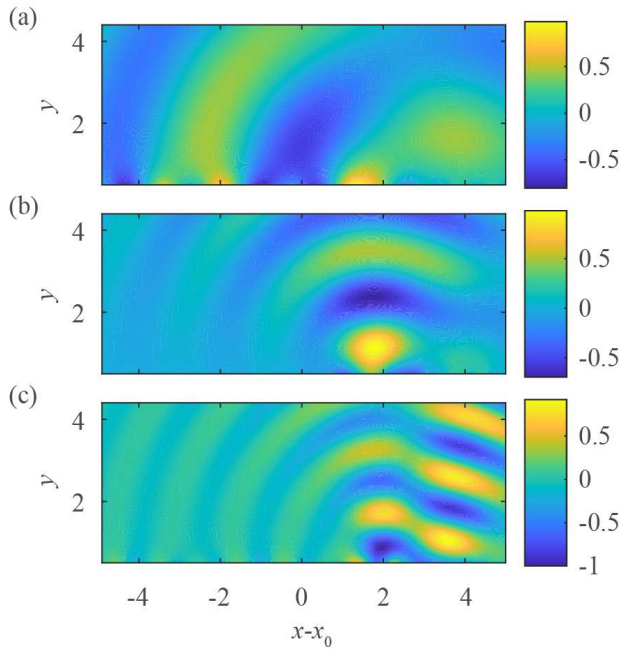}
        \caption{\addd{The \addd{normalized} near-field pressure immediately outside the jet~($x_{0}$ is the starting position of the effective source) due to the  one-cell interaction~($\alpha_{\mathrm{i}}\neq 0$). The Mach number of the fully-expanded jet flow is 1.5. The shock spacing is 2.236, therefore the effective sound source is located between 0 and 2.236. The labels (a), (b), and (c) represent the results at the fundamental frequency, its first and second harmonics, respectively.}}
    \label{fig:spl_1.5}
\end{figure}
	\begin{figure}
    \centering
    \includegraphics[width = 0.6\textwidth]{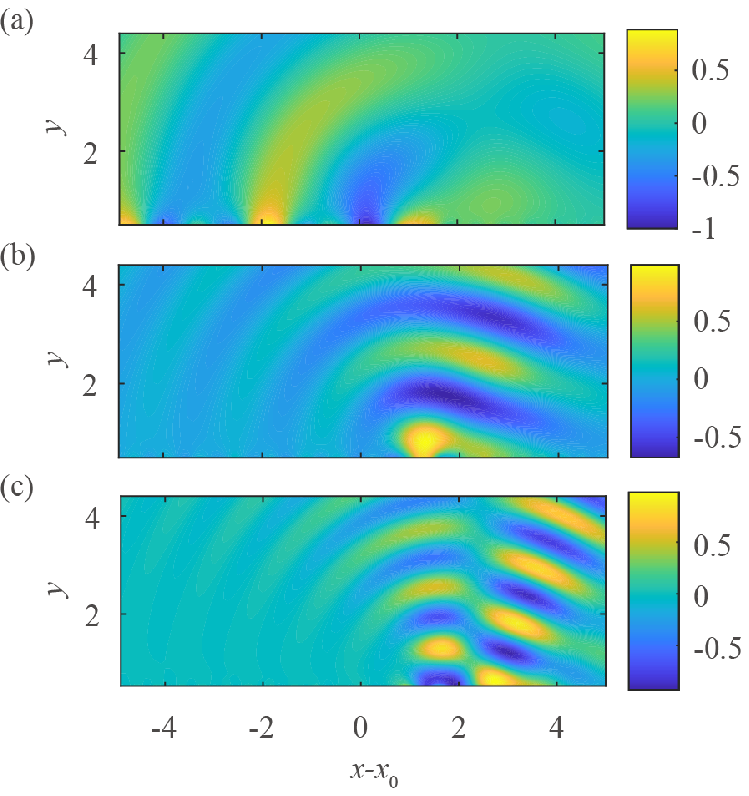}
    \caption{\addd{The \addd{normalized} near-field pressure immediately outside the jet~($x_{0}$ is the starting position of the effective source) due to the  one-cell interaction~($\alpha_{\mathrm{i}}\neq 0$). The Mach number of the fully-expanded jet flow is 1.3. The shock spacing is 1.661, therefore the effective sound source is located between 0 and 1.661. The labels (a), (b), and (c) represent the results at the fundamental frequency, its first  and second harmonics, respectively.}}
    \label{fig:spl_1.3}
\end{figure}
	\begin{figure}
    \centering
    \includegraphics[width = 0.6\textwidth]{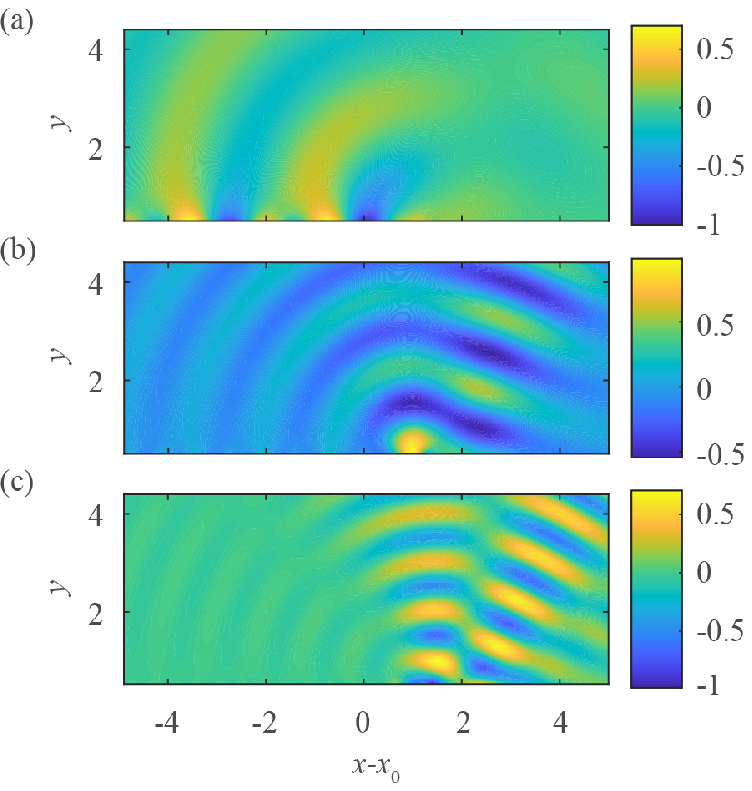}
    \caption{\addd{The \addd{normalized} near-field pressure immediately outside the jet~($x_{0}$ is the starting position of the effective source) due to the  one-cell interaction~($\alpha_{\mathrm{i}}\neq 0$). The Mach number of the fully-expanded jet flow is 1.2. The shock spacing is 1.327, therefore the effective sound source is located between 0 and 1.327. The labels (a), (b), and (c) represent the results at the fundamental frequency, its first harmonic and second harmonics, respectively.}}
    \label{fig:spl_1.2}
\end{figure}
 To further examine the noise generation due to the interaction between shock and instability waves, we can calculate the near-field pressure perturbation \adddd{$p_{+}$} by numerically integrating~(\ref{equ:numerical integration}). In the results shown below, all lengths are nondimensionalized by the height of the jet $D$. The shock spacing can be obtained from~(\ref{equ:shock space}). Figure~\ref{fig:spl_1.5} shows the pressure perturbation immediately outside the jet due to the interaction between instability waves and one shock cell. Labels (a), (b), and (c) represent the results at the fundamental frequency, its first and second harmonics, respectively.
It can be seen that at the fundamental frequency, the near-field pressure has a dominated distribution in the upstream direction, while at its first harmonic it shows a strong distribution perpendicular to the jet flow. This is in good agreement with the far-field directivity pattern shown in figure~\ref{n=1_consider}. For the second harmonic, two major distribution lobes are visible around $80^\circ$ and $110^\circ$, whereas the radiation at $\theta=90^\circ$ is relatively weak. These results are in good agreement with the directivity patterns in the far field, as shown in figure~\ref{n=1_consider}(1). When the Mach number of the fully-expanded jet changes to 1.3 or 1.2, as shown in figures~\ref{fig:spl_1.3} and~\ref{fig:spl_1.2}, respectively, similar agreement between the near- and far-field is achieved, and we omit a repetitive description for brevity.

It is interesting to note that for the fundamental frequency the near-field pressure perturbations span the entire shock cell, while for the harmonics they appear to be somewhat localized near the end of the shock, as illustrated in figures~\ref{fig:spl_1.5}(b-c),~\ref{fig:spl_1.3}(b-c), and~\ref{fig:spl_1.2}(b-c). This phenomenon was also  reported in a recent experiment conducted by~\citet{2020_sound_sources}. \adddd{However, since the present model has a periodic nature by construction, we cannot determine whether the sources are  physically distributed or localized. But it is interesting to note these near-field behaviours.}
\begin{figure}
    \centering
    \includegraphics[width = 0.55\textwidth]{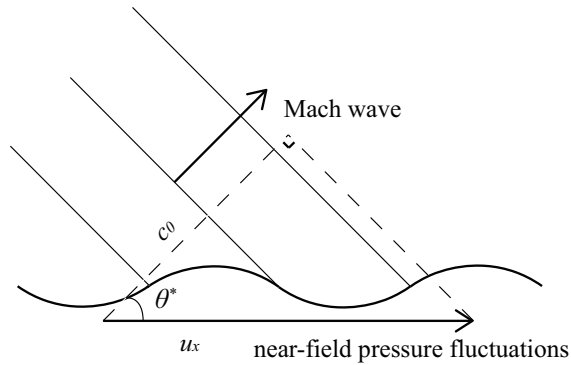}
    \caption{Schematic of the Mach wave radiation in supersonic jets. The phase velocity of the near-field pressure fluctuations along the jet flow is $u_{x}$, while the phase velocity of the radiated sound is $c_{0}$.}
    \label{fig:machwave}
\end{figure}

\begin{figure}
    \centering
   \includegraphics[width = 0.55\textwidth]{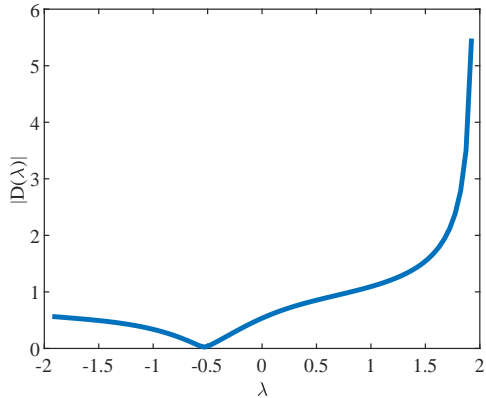}
   \caption{\addd{$|D(\lambda)|$ at the frequency of the fundamental tone. The Mach number of the fully-expanded jet flow is 1.3.}}
    \label{fig:mechanism_3}
\end{figure}

Using the model and results of the near-field pressure fluctuations, we now are in a position to examine the noise generation mechanism due to the interaction between shock and instability waves. As illustrated in section~\ref{subsec:sound}, the velocity potential takes the form
\begin{equation}
    \phi_{i+}=  \frac{U D}{\pi}\int_{-\infty}^{+\infty}D_{1}(\lambda)\mathrm{e}^{-(\mathrm{i}\lambda x+\mathrm{i}\omega t+\gamma_{+}y)}\mathrm{d}\lambda\mathrm{e}^{-\mathrm{i}\omega t}.
    \label{equ:3.8}
\end{equation}
In the far field,  $\phi_{i+}$ can be estimated by the saddle point method. The saddle point $\lambda_{0}$ is $-\omega M_{+}\cos\theta$, where $\theta$ represents the observer angle to the downstream direction. It can be shown that the $D_{1}(\lambda)$ is connected with the Fourier \add{transform} of the near-field pressure fluctuations~(along a fixed $y_{0}$). It is known that in supersonic jets, the phase velocity of the near-field pressure fluctuations along the jet flow can be supersonic relative to the ambient speed of sound, which leads to the Mach
wave radiation~\citep{tam_machwave}. As shown in figure~\ref{fig:machwave}, the Mach angle satisfies
\begin{equation}
    \theta^{*}=\arccos(\frac{c_{0}}{u_{x}}),
    \label{equ:mach wave}
\end{equation}  
where $\theta^{*}$ represents the direction of Mach wave radiation, and $u_{x}$, $c_{0}$ denote the phase velocities along the jet flow and the radiation direction, respectively. In our case, the phase velocity of the near-field pressure fluctuations in the $+x$ direction is equal to $-\omega/\lambda$, while the Mach wave has the phase velocity of $1/M_{+}$ in the radiation direction. So from~(\ref{equ:mach wave}), we can obtain 
\begin{equation}
    \theta^{*}=\arccos (-\frac{\lambda}{M_{+}\omega}).
    \label{3.10:equ}
\end{equation}
It is straightforward to find that $\lambda=-\omega M_{+}\cos\theta^{*}$, which is exactly the same as the saddle point $\lambda_{0}$. In fact, the saddle point precisely matches the $x$ component of the wavenumber of the Mach wave propagating to the $\theta$ direction. As the observer angle~(or the Mach wave radiation angle) changes from $0$ to $\pi$, the saddle point $\lambda_{0}$ changes from $-\omega M_{+}$ to $\omega M_{+}$ correspondingly. Therefore, the noise radiated at angle $\theta$ is directly related to $D_{1}(\lambda_{0})$ through the Mach wave mechanism, as shown in~(\ref{equ:3.8}). We may therefore examine the directivities of the sound generation by examining $|D_{1}(\lambda)|$ between $-\omega M_{+}$ and $\omega M_{+}$. Figure~\ref{fig:mechanism_3} shows $|D_{1}(\lambda)|$ as a function of $\lambda$. We see that sound radiates primarily to the upstream direction~($\lambda>0$), which is in good agreement with the experimental data.

 However, compared with the result in figure~\ref{n=1_consider}, there is no quick decay of $D_{1}(\lambda)$ as $\lambda\rightarrow\omega M_{+}$, so the sound at $180^\circ$ appears to be the strongest. This appears to contradict figure~\ref{n=1_consider}. This is because $|D_{1}(\lambda_{0})|$ can only show the overall shape of the directivity pattern, but the directivity is in fact determined by $|D_{1}(\lambda_{0})\sin\theta|$ instead. We can show this by rewriting~(\ref{equ:3.8}) as 
\begin{equation}
    \phi_{i+}=  \frac{U D}{\pi}\int_{-\infty}^{+\infty}D_{1}(\lambda)\mathrm{e}^{-r(\mathrm{i}\lambda \cos\theta+\gamma_{+}\sin\theta)}\mathrm{d}\lambda\mathrm{e}^{-\mathrm{i}\omega t},
    \label{equ:whatever}
\end{equation}
where $r$ again denotes the radial distance, and $\theta$ represents the observer angle. We consider the substitution
\begin{equation}
    \lambda=-\omega M_{+}\cos(\theta+\beta),
\end{equation}
with which~(\ref{equ:whatever}) can be rewritten as
\begin{equation}
      \phi_{i+}=  \frac{U D}{\pi}\int_{P^{*}}D_{1}(-\omega M_{+}\cos(\theta+\beta))\sin(\theta+\beta)\mathrm{e}^{\mathrm{i}\omega M_{+} r \cos\beta}\mathrm{d}\beta\mathrm{e}^{-\mathrm{i}\omega t},
      \label{equ:3.13}
\end{equation}
where $P^{*}$ denotes the new integration contour in the complex $\beta$ plane. We see that the exponent term in~(\ref{equ:3.13}) is independent of $\theta$, and now if we use the saddle point method~(the saddle point is $\beta_{0}=0$), we obtain
\begin{equation}
      \phi_{i+}=  \frac{U D}{\pi}D_{1}(\lambda_{0})\sin\theta F(r,\beta),
      \label{equ:3.14}
\end{equation}
where $F(r,\beta)$ is independent of $\theta$.
So the result which determines the far-field directivity pattern is indeed $|D_{1}(\lambda_{0})\sin\theta|$.

\begin{figure}
    \centering
    \includegraphics[width = 0.3\textwidth]{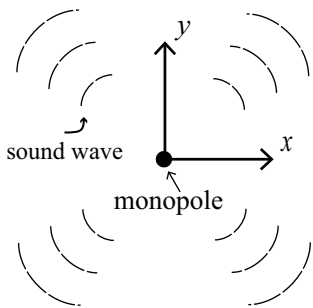}
    \caption{The schematic of a monopole located at $(0,0)$ and its radiated sound.}
    \label{fig:monopole}
\end{figure}

To better understand this, we take the classical sound field of a monopole as an example. Consider a monopole located at $(0,0)$ in the $x-y$ plane, as shown in figure~\ref{fig:monopole}. Its velocity potential $\phi^{*}$ satisfies the homogeneous Helmholtz equation, i.e.
\begin{equation}
    \bnabla^{2}\phi^{*}+\omega^{2}M_{+}^{2}\phi^{*}=0.
    \label{equ:helmholze}
\end{equation}
Its axisymmetric solution in the infinite space takes the form $\phi^{*}=H_{0}^{1}(\omega M_{+}r)$, where $H^{1}_{0}(\omega M_{+}r)$ is the $0$th-order Hankel function of the first kind and $r$ is the radial distance to the sound source. Note such a solution has a uniform directivity. Although we have obtained the exact solution, we still repeat the calculation process to demonstrate how the factor $\sin\theta$ emerges. 
 Similar to~(\ref{equ:numerical integration}), Fourier \add{transform} can be used to solve~(\ref{equ:helmholze}), and the solution is
 \begin{equation}
        \phi^{*}(x,y)=\frac{1}{2\pi}\int_{-\infty}^{+\infty}D^{*}(\lambda)\mathrm{e}^{-(\mathrm{i}\lambda x+\gamma_{+}y)}\mathrm{d}\lambda.
        \label{equ:numerical integration_2}
\end{equation}
Since $\phi^{*}$ is already known, $D^{*}(\lambda)$ can be obtained by taking the Fourier \add{transform} of $H_{0}^{1}(\omega M_{+} r)$ along $y=0$. It is straightforward to find that
\begin{equation}
    D^{*}(\lambda)=\dfrac{2}{\sqrt{\omega^{2}M_{+}^{2}-\lambda^{2}}},
    \label{equ:d_lambda}
\end{equation}
where suitable branch of $\sqrt{\omega^{2}M_{+}^{2}-\lambda^{2}}$ is chosen. In the far field, the saddle point method can be used to estimate~(\ref{equ:numerical integration_2}). The saddle point is $\lambda_{0}$, and the final result reduces to
    \begin{equation}
        \phi^{*}(r,\theta)=\frac{\sqrt{M_{+}\omega}}{\sqrt{2\pi}}D^{*}(\lambda_{0}){\rm s i n} \theta \frac{\mathrm{e}^{\mathrm{i}\omega(M_{+}r-\pi/4)}}{\sqrt{r}}+O(r^{-3/2}).
        \label{steepest_decent_way_D_star}
    \end{equation}
Here $D^{*}(\lambda_{0})=2/\omega M_{+}\sin\theta$, and $\phi^{*}(r,\theta)=\sqrt{2/\pi\omega M_{+}r}\mathrm{e}^{\mathrm{i}\omega(M_{+}r-\pi/4)}$, which is exactly the far-field approximation of $H_{0}^{1}(\omega M_{+} r)$. We can see that the directivity is indeed determined by $|D^{*}(\lambda_{0})\sin\theta|$ rather than $|D^{*}(\lambda_{0})|$. In fact, the coefficient $D^{*}({\lambda}_{0})=2/\omega M_{+}\sin\theta\rightarrow \infty$  as the observer angle $\theta\rightarrow 180^\circ$. It is $|D^{*}(\lambda_{0})\sin\theta|$ that remains bounded and is independent of $\theta$ as expected from the solution $H^{1}_{0}(\omega M_{+} r).$

 Figure~\ref{fig:mechanism_3} only shows results for the fundamental frequency at $M_{-}=1.3$. Similar results can be obtained for higher harmonics at other Mach numbers. We see that noise is primarily generated through the Mach wave mechanism.
  \add{Note that Mach wave radiation can also occur in perfectly-expanded supersonic jets via jet instability waves. However, in this paper, we focus on the interaction between shock and instability waves, where the near-field fluctuations with supersonic phase speed lead to sound generation. This is not to be confused with the convectional Mach wave radiation due to the jet instability waves in  perfectly-expanded supersonic jets.}

\section{Conclusion}
\label{section:conclusion}
An analytical model is developed in this paper to predict the sound arising from the interaction between shock and instability waves in imperfectly-expanded 2D jets. Both shock and instability waves are assumed to be of small amplitudes so that linear theories may be used. A vortex-sheet model is used to describe the base jet flow, and 2D Euler equations are subsequently linearised around this base flow to determine the governing equations for shock, instability waves and their interaction, respectively. The interaction between shock and instability waves is determined by solving an inhomogeneous wave equation while simultaneously matching kinematic and dynamic conditions on the vortex sheets. The generated sound in the far field is obtained in a closed form after Fourier \add{transform} is used in conjunction with the saddle point method. 

The screech frequencies are determined by using the constructive
interference assumption proposed by Powell and show good agreement with experimental results. The model can be used to predict the sound due to the instability waves interacting with one shock cell, as well as that with a number of shock cells. \adddd{The directivity of the sound due to the one-cell interaction is shown to resemble that of the total sound field. It is interesting to note that the noise directivity is sensitive to the local growth rate of the instability waves interacting with the shock cells to generate sound and may be used to explain the discrepancies observed across different experiments. When multiple shock cells are included, the present model shows better agreement with experiments and simulations than the monopole array theory. In particular, the present model corrextly captures the rapid decay of the acoustic radiation when $\theta$ approaches $180^\circ$ and $0^\circ$, respectively.} In particular, noise radiation primarily occurs in the upstream direction but becomes weaker as the observer angle gradually approaches 180 degrees, which is in better agreement with experimental results compared with earlier models. 

The near-field pressure fluctuation due to the shock-instability interaction is subsequently studied. It is shown that the near-field pressure fluctuation has a distribution that is consistent with the far-field directivity patterns. By examining the wavenumber matching of the near-field pressure, we find that noise is generated primarily through the Mach wave mechanism. It is shown that the model developed in this paper can correctly capture the essential physics and may be used to further study the screech in imperfectly-expanded supersonic jets.
\section*{Acknowledgments}
The authors gratefully acknowledge the funding under Marine S\&T Fund of Shandong Province for Pilot National Laboratory for Marine Science and Technology (Qiangdao) (No. 2022QNLM010201). The second author (B.L.) thanks Professor A. P. Dowling for an earlier stimulating discussion on jet instability waves.

Declaration of Interests. The authors report no conflict of interest.
\appendix
\section{}\label{appA}
The corresponding velocity, pressure perturbations within the jet, and the vortex  are
\begin{equation}
    u_{m}=a A \cos(a \beta y)\cos(a x),
\end{equation}
\begin{equation}
    v_{m}=-a\beta A \sin(a \beta y)\sin(a x),
\end{equation}
\begin{equation}
    p_{m}=-\rho_{-}a U  A \cos(a \beta y)\cos(a x).
\end{equation}
Note the subscripts in parameters $A_{1}$ and $a_{1}$ are omitted for
clarity. It is worth noting that all these shock-associated perturbations are independent of time. The vortex sheet deflection at the boundary of jet flow reduces to
\begin{equation}
  {h}_{m}= 
    \begin{cases}
      \dfrac{A}{U}\beta\sin(\dfrac{1}{2}a\beta)\cos(a x), & y=1/2 \\
      -\dfrac{A}{U}\beta\sin(\dfrac{1}{2}a\beta)\cos(a x), & y=-1/2.        
  \end{cases}
\end{equation}

\section{}\label{appB}
 The corresponding pressure and velocity are
 \begin{equation}
     p_{v}=\mathrm{i}\omega U^{2} \mathrm{e}^{\mathrm{i}(\alpha x-\omega t)}\times
     \begin{cases}
        \dfrac{\rho_{+}}{M_{+}^{2}}\mathrm{e}^{-m_{+}y},& y>\frac{1}{2}\\
        \dfrac{\rho_{-}}{M_{-}^{2}}(k_{2}\mathrm{e}^{-m_{-}y}+k_{1}\mathrm{e}^{m_{-}y}),&y\leq |\frac{1}{2}|\\
        \dfrac{\rho_{+}}{M_{+}^{2}}k_{3}\mathrm{e}^{m_{+}y},&y<-\frac{1}{2},
     \end{cases}
 \end{equation}
 \begin{equation}
     u_{v}=\mathrm{i}\alpha U\mathrm{e}^{\mathrm{i}(\alpha x-\omega t)}\times
     \begin{cases}
         \dfrac{1}{M_{+}^{2}}\mathrm{e}^{-m_{+}y},&y>\frac{1}{2}\\
         \dfrac{1}{M_{-}^{2}}\dfrac{\omega}{\omega-\alpha}(k_{2}\mathrm{e}^{-m_{-}y}+k_{1}\mathrm{e}^{m_{-}y}),&y\leq |\frac{1}{2}|\\
         \dfrac{1}{M_{+}^{2}}k_{3}\mathrm{e}^{m_{+}y},&y<-\frac{1}{2},
     \end{cases}
 \end{equation}
 \begin{equation}
     v_{v}=U \mathrm{e}^{\mathrm{i}(\alpha x-\omega t)}\times
     \begin{cases}
         \dfrac{1}{M_{+}^{2}}m_{+}\mathrm{e}^{-m_{+}y},&y>\frac{1}{2}\\
         \dfrac{1}{M_{-}^{2}}\dfrac{\omega}{\omega-\alpha}m_{-}(-k_{2}\mathrm{e}^{-m_{-}y}+k_{1}\mathrm{e}^{m_{-}y}),&y\leq|\frac{1}{2}|\\
         \dfrac{1}{M_{+}^{2}}m_{+}k_{3}\mathrm{e}^{m_{+}y},&y<-\frac{1}{2}.
     \end{cases}
 \end{equation}
The deflection of the jet boundary due to the instability waves is
\begin{equation}
  {h}_{v}(x,t)= 
    \begin{cases}{}
      -\dfrac{\mathrm{i}}{\omega}\dfrac{m_{+}}{M_{+}^{2}}\mathrm{e}^{-\frac{1}{2}m_{+}}\mathrm{e}^{\mathrm{i}(\alpha x-
    \omega t)},  &y=1/2 \\
     \dfrac{\mathrm{i}}{\omega}k_{3}\dfrac{m_{+}}{M_{+}^{2}}\mathrm{e}^{-\frac{1}{2}m_{+}}\mathrm{e}^{\mathrm{i}(\alpha x-\omega t)}, & y=-1/2.
  \end{cases}
\end{equation}
\bibliographystyle{jfm}
\bibliography{jfm-instructions}
\end{document}